\documentclass[pdflatex,sn-mathphys-ay]{sn-jnl}

\usepackage{graphicx}%
\usepackage{amsmath,amssymb,amsfonts}%
\usepackage{amsthm}%
\usepackage{xcolor}% Required only while author comments remain
\usepackage{booktabs}%
\usepackage{subcaption}
\usepackage{microtype}
\usepackage{siunitx}
\usepackage{bbm}
\usepackage{csquotes}
\usepackage{longtable}

\graphicspath{{./art/}}

\theoremstyle{thmstyleone}%
\newtheorem{theorem}{Theorem}%
\newtheorem{lemma}{Lemma}%

\theoremstyle{thmstyletwo}%
\newtheorem{assumption}{Assumption}%

\makeatletter
\AtBeginDocument{%
  \let\snjnl@proof\proof
  \let\snjnl@endproof\endproof
  \renewenvironment{proof}[1][]{%
    \if\relax\detokenize{#1}\relax
      \snjnl@proof
    \else
      \snjnl@proof[\proofname\ \ignorespaces#1]%
    \fi
  }{\snjnl@endproof}%
}
\makeatother

\newcommand{\ind}{\mathbbm{1}}
\newcommand{\prob}{\operatorname{pr}}

\newcommand{\Rset}{\mathbb{R}}
\newcommand{\bX}{X}
\newcommand{\bZ}{Z}
\newcommand{\ba}{a}

\newcommand{\bD}{D}
\newcommand{\btheta}{\theta}
\newcommand{\bTheta}{\Theta}

\newcommand{\clr}{\operatorname{clr}}

\begin{document}

\title{Clustering extremal dependence structures via Wasserstein distances between angular measures}

% =============================================================================
% Authors and affiliations
% =============================================================================

\author*[1]{\fnm{K.} \sur{Gasser}}\email{kamal.gasser@uclouvain.be}

\author[2,1]{\fnm{J.} \sur{Segers}}\email{jjjsegers@kuleuven.be}

\author[3]{\fnm{F.} \sur{Ragone}}\email{fr120@le.ac.uk}

\affil[1]{\orgdiv{Institute of Statistics, Biostatistics and Actuarial Sciences, LIDAM}, \orgname{UCLouvain}, \orgaddress{\street{Voie du Roman Pays 20}, \postcode{B-1348} \city{Louvain-la-Neuve}, \country{Belgium}}}

\affil[2]{\orgdiv{Department of Mathematics}, \orgname{KU Leuven}, \orgaddress{\street{Celestijnenlaan 200B}, \postcode{B-3001} \city{Heverlee}, \country{Belgium}}}

\affil[3]{\orgdiv{School of Computing and Mathematical Sciences}, \orgname{University of Leicester}, \orgaddress{\street{University Road}, \city{Leicester} \postcode{LE1 7RH}, \country{U.K}}}

% =============================================================================
% Abstract and keywords
% =============================================================================

\abstract{Grouping random vectors by their joint extremal behaviour is a recurring task in climate science, hydrology, and finance. Classical clustering is driven by typical observations, while existing extremal procedures either summarize dependence by a single coefficient or discretize the simplex in an ad hoc way. We propose a dissimilarity that compares complete angular measures by optimal transport: each angular measure is estimated above a high radial threshold, mapped from the simplex into a Euclidean space of Aitchison coordinates, and compared by the 1-Wasserstein distance. The methodology is fully nonparametric and requires no fitted model. The dissimilarities feed any distance-based clustering algorithm. Under intermediate-threshold asymptotics and mild moment and separation conditions, hierarchical clustering and partitioning around medoids based on them recovers the latent partition with probability tending to one. In simulations, the proposed approach matches established competitors when clusters differ mainly in the strength of extremal dependence and performs substantially better when equal-strength clusters differ in the shape of their angular measures, especially when asymmetry drives the separation. Applications to rain--snowmelt--runoff extremes in Swiss catchments and to extremal co-movement between international equity markets and the US market reveal hydrologically and economically interpretable groups.}

\keywords{Aitchison coordinate, Angular measure, Compound extreme, Extreme value theory, Optimal transport, Wasserstein distance}

\maketitle

% =============================================================================
% Main paper
% =============================================================================

\section{Introduction}\label{sec:Intro}
Clustering is one of the fundamental tasks in unsupervised statistical learning, aiming to partition vector of variables into homogeneous groups while maximizing their separation from other groups. As such, it has become a tool across a wide range of scientific disciplines, including biology, economics, finance, environmental sciences and climatology, where it is routinely used to uncover hidden structures.
In many applications, the primary scientific interest instead lies in understanding the behaviour of rare, high-impact events. In climate science, for example, identifying regions that exhibit similar patterns of joint heatwaves and droughts is of direct interest \citep{zscheischler2020typology}; in hydrology, clustering river basins according to the dependence structure of flood characteristics can support hazard mapping; and in finance, grouping markets according to their tendency to experience simultaneous crashes provides valuable information for systemic risk management and portfolio diversification. These applications call for clustering methods that focus specifically on the extremal behaviour of multivariate random vectors.

The development of clustering methods specifically designed for extreme events has attracted increasing attention over the last decade. Early contributions focused on the regionalization of spatial extremes by adapting classical clustering algorithms to dependence measures arising from extreme-value theory. \citet{bernard2013clustering} proposed a partitioning around medoids algorithm in which the classical Euclidean distance is replaced by the $F$-madogram, yielding clusters of sites exhibiting similar extremal dependence while preserving a meaningful representative within each cluster. More recently, \citet{kiriliouk2020climate} extended this line of research by proposing a regionalization approach based on the tail dependence coefficient $\chi$, providing an alternative characterization of extremal dependence between locations. In parallel, \citet{janssen2020kmeans} introduced an extremal version of spherical $k$-means, where clustering is performed directly on extremal observations to estimate representative directions of extremal dependence, together with consistency guarantees for the resulting cluster centres. Beyond these approaches based on extreme-value theory, \citet{benevento2025comonotonic,benevento2024tail} developed several copula-based clustering procedures combining tail dependence measures and spatial information to construct dissimilarity matrices, with applications to environmental and climatological data.

More recently, the focus has shifted from clustering spatial locations to clustering multivariate extremal dependence structures themselves. \citet{fuchs2021dissimilarity} proposed a dependence-based clustering framework relying on the notion of $\pi$-comonotonicity, allowing multivariate random vectors to be grouped according to their copula. Although broadly applicable, this framework is designed for general dependence structures rather than specifically for extremal dependence, and therefore does not fully exploit the asymptotic theory of multivariate extremes. From an extreme-value perspective, \citet{boulin2025high} introduced the asymptotic-independence block model, where variables are partitioned into groups that are asymptotically independent. Under this model, clusters are estimated through a statistically consistent procedure with polynomial computational complexity. While elegant from a theoretical viewpoint, the asymptotic-independence block assumption may be restrictive in practice, as extremal dependence structures often exhibit more complex interactions than strict block-wise asymptotic independence.

\citet{vignotto2021clustering} proposed a fully nonparametric clustering approach based on partitioning the extreme region and comparing the resulting empirical probabilities through a Kullback--Leibler dissimilarity. More recently, \citet{engelke2026kullback} established asymptotic theory for the underlying Kullback--Leibler statistic in a two-sample testing framework. Although this construction avoids restrictive parametric assumptions, its implementation requires a predefined partition, whose choice may substantially influence the resulting dissimilarities and clustering. \citet{fomichov2023spherical} developed spherical \(k\)-principal-components clustering to detect groups of concomitant extremes. More recently, \citet{o2025clustering} introduced a conditional extremes clustering framework, which compares extremal dependence through fitted conditional extremes models. By relying on the conditional extremes framework, the method provides a flexible characterization of extremal dependence and has demonstrated excellent empirical performance. Nevertheless, its implementation requires fitting a conditional model for each random vector, which can become computationally demanding for large collections of datasets. Moreover, the method typically relies on Gaussian residual assumptions, which may be inadequate.

We propose an optimal-transport framework for clustering multivariate extremal dependence structures. The approach builds on the Wasserstein--Aitchison construction of \citet{lhaut2026simulation}, who map angular observations from the simplex to a linear space using Aitchison coordinates and train a Wasserstein generative adversarial network to learn the angular measure and generate multivariate extreme observations. We use this construction for a different purpose: Wasserstein distances are computed directly between empirical angular measures in Aitchison coordinates, producing a dissimilarity metric that can be supplied to any distance-based clustering algorithm. The resulting procedure compares complete extremal dependence structures without summary coefficients, ad hoc discretization of the simplex, or fitted parametric or generative models.

We establish convergence of the empirical dissimilarities and consistency of the resulting clustering procedures. An extensive simulation study compares the proposed approach with existing methods across a wide range of extremal dependence structures. Applications to climatological and financial data illustrate its practical usefulness.

\section{Methods}\label{sec:Methods}
\subsection{Statistical framework}\label{sec:Framework}

We consider $n$ random vectors of dimension $p$, each observed through $T$ independent and identically distributed realizations. For $i=1,\ldots,n$ and $t=1,\ldots,T$, let
\[
\bX_{i,t}
=
\bigl(
X_{i,t}^{(1)},
\ldots,
X_{i,t}^{(p)}
\bigr)
\in \Rset^p.
\]
We assume that $\bX_{i,1},\ldots,\bX_{i,T}$ have the same distribution function $F_i$. Moreover, we assume that the marginal distributions are continuous, and that the $n$ random vectors are partitioned into $K$ latent groups. Let $g(i)\in\{1,\ldots,K\}$ denote the group membership of vector $i$. Then random vectors belonging to the same group are assumed to share a common dependence structure. To lighten the notation, we write $\bX_i=\bigl(X_i^{(1)},\ldots,X_i^{(p)}\bigr)$ for a generic random vector with distribution function $F_i$, that is, for a random vector distributed as any of $\bX_{i,1},\ldots,\bX_{i,T}$. For each $i$, define the probability integral transforms
\[
U_i^{(j)}
=
F_i^{(j)}
\bigl(
X_i^{(j)}
\bigr),
\qquad j=1,\ldots,p,
\]
where $F_i^{(j)}$ is the marginal distribution function of the $j$th component of random vector $i$.

By Sklar's theorem, the joint distribution of the $p$-variate random vector can be represented through a copula \citep{sklar1959fonctions}. Specifically,
\[
\prob\!\bigl(
X_i^{(1)}\leq x_1,\ldots,
X_i^{(p)}\leq x_p
\bigr)
=
C_{i}
\!\bigl(
F_i^{(1)}(x_1),\ldots,
F_i^{(p)}(x_p)
\bigr),
\]
where $C_{i}$ denotes the copula associated with random vector $i$. We assume that each $C_{i}$ is in the domain of attraction of an extreme-value copula, which guarantees the existence of an angular measure. Background on the extreme-value copulas used in the simulation study is given in Section~\ref{sec:ev_copulas} of the Supplementary Material.

We assume that random vectors belonging to the same latent group share a common angular measure $H_{g(i)}$ concentrated on the open simplex $\Delta_{p-1}^{\circ}$, as described in Section~\ref{sec:angular_measure}, where $g:\{1,\ldots,n\}\to \{1,\ldots,K\}$ denotes the group-assignment function. The objective is to construct a dissimilarity measure between the $n$ angular measures, which can then be used as input to a clustering algorithm in order to partition the $n$ random vectors into the $K$ latent groups according to their extremal dependence structure.

\subsection{Angular measure}\label{sec:angular_measure}

To characterize extremal dependence, we rely on the classical radial--angular decomposition from multivariate extreme-value theory \citep{beirlant2004statistics,resnick2007heavy}. Following the marginal standardization introduced in the previous section, each component is transformed to a standard Pareto scale. For $j=1,\ldots,p$, define
\begin{equation}
Z_i^{(j)}
=
\frac{1}{1-U_i^{(j)}}
=
\frac{1}{1-F_i^{(j)}\!\bigl(X_i^{(j)}\bigr)}.
\label{eq:pareto_transformation}
\end{equation}
By construction, $\prob\!\bigl(Z_i^{(j)}>z\bigr)=z^{-1}$ for $z\geq 1$, so that each transformed margin follows a standard Pareto distribution. We therefore obtain the transformed random vector $\bZ_i=\bigl(Z_i^{(1)},\ldots,Z_i^{(p)}\bigr)$.

Using the $L_1$-norm, the radial and angular components associated with $\bZ_i$ are defined by
\begin{equation}
R_i
=
\left\|\bZ_i\right\|_1
=
\sum_{j=1}^{p}Z_i^{(j)},
\qquad
\bTheta_i
=
\frac{\bZ_i}{R_i}.
\label{eq:radial_angular_component}
\end{equation}
By construction, $\sum_{j=1}^{p}\Theta_i^{(j)}=1$, so that $\bTheta_i$ belongs to the unit simplex
\begin{equation}
\Delta_{p-1}
=
\bigl\{
\bTheta\in[0,1]^p:
\sum_{j=1}^{p}\Theta^{(j)}=1
\bigr\}.
\label{eq:unit_simplex}
\end{equation}
The radial variable $R_i$ measures the overall magnitude of an observation, whereas the angular component $\bTheta_i$ describes how this magnitude is distributed across the different variables. Under standard multivariate regular variation assumptions, the conditional distribution satisfies
\begin{equation}
\prob(
\bTheta_i\in A
\mid
R_i>r
)
\longrightarrow
H_i(A),
\qquad r\to\infty,
\label{eq:angular_measure_limit}
\end{equation}
for Borel sets $A\subset\Delta_{p-1}$ such that $H_i(\partial A)=0$, where $\partial A$ denotes the boundary of $A$ relative to $\Delta_{p-1}$. The limiting probability measure $H_i$ is called the angular measure and characterizes the extremal dependence structure of random vector $i$ \citep{resnick2007heavy,de2006extreme}.
The angular measure $H_i$ may assign positive probability to the boundary of the simplex $\Delta_{p-1}$, namely to subsets where at least one coordinate is equal to zero. From the perspective of multivariate extremes, this represents situations in which some components of $\bX$ are extreme while others remain non-extreme. Such configurations, often referred to as extreme directions \citep{mourahib2025multivariate}, are excluded from our methodology. We assume that the angular measure concentrates on the open simplex
\begin{equation}
\Delta_{p-1}^{\circ}
=
\bigl\{
\bTheta\in(0,1)^p:
\sum_{j=1}^{p}\Theta^{(j)}=1
\bigr\}.
\label{eq:open_unit_simplex}
\end{equation}

The geometry of the angular measure reflects the strength and structure of extremal dependence. When extremes tend to occur in only a few components, the angular mass concentrates near the vertices of the simplex. In contrast, when large values occur simultaneously across several components, the angular mass tends to concentrate in the interior of the simplex \citep{coles2001introduction}. Figure~\ref{fig:angular_measure_bivariate}  of the Supplementary Material illustrates this behaviour for a bivariate random vector under two extremal dependence regimes. In the low-dependence regime, the angular observations concentrate near the vertices of the simplex, indicating that extremes mainly occur in one component at a time. In contrast, under strong extremal dependence, the angular observations concentrate near the centre of the simplex, reflecting the simultaneous occurrence of extremes.

In practice, the marginal distribution functions are unknown and the angular measure must be estimated from the observed sample. For $i=1,\ldots,n$, $t=1,\ldots,T$, and $j=1,\ldots,p$, define the empirical probability integral transform
\begin{equation}
\widehat U_{i,t}^{(j)}
=
\frac{\operatorname{rank}\bigl(X_{i,t}^{(j)}\bigr)}{T+1},
\label{eq:empirical_uniform_transformation}
\end{equation}
where, the rank of $X_{i,t}^{(j)}$ is computed among $\{X_{i,s}^{(j)}: s=1,\ldots,T\}$. The corresponding empirical standard Pareto transformation is
\begin{equation}
\widehat Z_{i,t}^{(j)}
=
\frac{1}{1-\widehat U_{i,t}^{(j)}}.
\label{eq:empirical_pareto_transformation}
\end{equation}
Write $\widehat{\bZ}_{i,t}=\bigl(\widehat Z_{i,t}^{(1)},\ldots,\widehat Z_{i,t}^{(p)}\bigr)$. The empirical radial and angular components are then defined by
\begin{equation}
\widehat R_{i,t}
=
\left\|\widehat{\bZ}_{i,t}\right\|_1
=
\sum_{j=1}^{p}\widehat Z_{i,t}^{(j)},
\qquad
\widehat{\bTheta}_{i,t}
=
\frac{\widehat{\bZ}_{i,t}}{\widehat R_{i,t}}.
\label{eq:empirical_angular_component}
\end{equation}
For a sufficiently high threshold $r_i$, let
\[
\mathcal T_i
=
\bigl\{
t\in\{1,\ldots,T\}:
\widehat R_{i,t}>r_i
\bigr\}
\]
denote the set of radial exceedances, and let $k_{i,T}=\lvert\mathcal T_i\rvert$ be the corresponding number of exceedances. The empirical angular measure associated with random vector $i$ is then defined, for every Borel set $A\subset\Delta_{p-1}$, by
\begin{equation}
\widehat H_{i,T}(A)
=
\frac{1}{k_{i,T}}
\sum_{t\in\mathcal T_i}
\ind\!\bigl\{
\widehat{\bTheta}_{i,t}\in A
\bigr\}.
\label{eq:empirical_angular_measure}
\end{equation}
Under standard multivariate regular variation assumptions, the empirical angular measure consistently estimates the underlying angular measure when the radial threshold is chosen according to an intermediate sequence. More precisely, let
\[
r_{i,T}
=
\frac{T}{\ell_{i,T}},
\qquad
\ell_{i,T}\to\infty,
\qquad
\frac{\ell_{i,T}}{T}\to 0,
\qquad
T\to\infty.
\]
Then $r_{i,T}\to\infty$, while the number of observations exceeding the threshold increases but remains asymptotically negligible relative to the sample size. Consequently, $\widehat H_i$ consistently estimates the limiting angular measure $H_i$. Consistency and asymptotic normality results for nonparametric estimators of spectral and angular measures have been established by \citet{einmahl2001nonparametric}, \citet{einmahl2009maximum}, \citet{janssen2020kmeans} and more recently, \citet{lhaut2024asymptotic}.

\subsection{Aitchison coordinates for angular measures}\label{sec:aitchison_coordinates}

The angular measure of a multivariate regularly varying random vector is supported on the unit simplex $\Delta_{p-1}$. Consequently, angular observations are compositional data, that is, vectors whose components are constrained to be positive and to sum to one. This simplex structure induces a non-Euclidean geometry and makes direct statistical analysis challenging \citep{aitchison1982statistical,egozcue2003isometric}. A central idea of compositional data analysis is to represent simplex-valued observations through coordinates in a linear Euclidean space. This is achieved using the centred log-ratio transformation
\begin{equation}
\clr(\btheta)
=
\left(
\log\frac{\theta^{(1)}}{f(\btheta)},
\ldots,
\log\frac{\theta^{(p)}}{f(\btheta)}
\right),
\qquad
\btheta\in\Delta_{p-1}^{\circ},
\label{eq:clr}
\end{equation}
where $f(\btheta)=\bigl(\prod_{j=1}^{p}\theta^{(j)}\bigr)^{1/p}$ denotes the geometric mean of the components of $\btheta$. The transformed vector belongs to the hyperplane
\[
\mathcal H
=
\left\{
w\in\Rset^p:
\sum_{j=1}^{p}w^{(j)}=0
\right\},
\]
which is a $(p-1)$-dimensional linear subspace of $\Rset^p$. The centred log-ratio transformation removes the simplex constraint and maps angular observations to an unconstrained Euclidean space while preserving their relative information. In particular, distances between compositions are naturally measured through the Aitchison distance
\begin{equation}
d_A(\btheta_1,\btheta_2)
=
\left\|
\clr(\btheta_1)-\clr(\btheta_2)
\right\|_2,
\label{eq:aitchison_distance}
\end{equation}
which corresponds to the Euclidean distance after the centred log-ratio transformation.

Following \citet{egozcue2003isometric}, we use the Helmert orthonormal basis of $\mathcal H$. For $j=1,\ldots,p-1$, define
\[
e_j
=
\left\{
\frac{j}{j+1}
\right\}^{1/2}
\left(
\underbrace{j^{-1},\ldots,j^{-1}}_{j\ {\rm entries}},
-1,
0,\ldots,0
\right).
\]
This basis is orthonormal in $\Rset^p$ and spans $\mathcal H$. We then represent each angular observation through its Aitchison coordinates. Let $\mathcal A:\Delta_{p-1}^{\circ}\to\Rset^{p-1}$ denote the Aitchison coordinate map, defined for every $\btheta\in\Delta_{p-1}^{\circ}$ by
\begin{equation}
\mathcal A(\btheta)
=
\left(
\left\langle\clr(\btheta),e_1\right\rangle,
\ldots,
\left\langle\clr(\btheta),e_{p-1}\right\rangle
\right).
\label{eq:aitchison_coordinates}
\end{equation}
Applying this map to the empirical angular observations in equation~\eqref{eq:empirical_angular_component} yields the Aitchison coordinates
\begin{equation}
\ba_{i,t}
=
\mathcal A\left(\widehat{\bTheta}_{i,t}\right)
\in\Rset^{p-1},
\qquad
t\in\mathcal T_i.
\label{eq:aitchison_observations}
\end{equation}
The use of Aitchison coordinates is particularly advantageous for clustering angular measures. Directly comparing distributions on the simplex may be difficult because the simplex is a constrained space where variability is compressed by the positivity and unit-sum constraints. The centred log-ratio transformation effectively decompresses the angular information by mapping the simplex onto a linear Euclidean space. As a consequence, differences between angular distributions become easier to detect and quantify. This idea was leveraged by \citet{lhaut2026simulation} to simulate multivariate extremes through generative adversarial networks operating in Aitchison coordinates.

In the proposed methodology, the empirical angular observations associated with each random vector, as defined in Eq.~\eqref{eq:empirical_angular_measure},  are first transformed into Aitchison coordinates, leading to an empirical probability measure on $\Rset^{p-1}$ defined, for every Borel set $B\subset\Rset^{p-1}$, in terms of the vectors $\ba_{i,t}$ in equation~\eqref{eq:aitchison_observations} by
\begin{equation}
\widehat G_{i,T}(B)
=
\frac{1}{k_{i,T}}
\sum_{t\in\mathcal T_i}
\ind\!\left\{
\ba_{i,t}\in B
\right\}.
\label{eq:empirical_aitchison_measure}
\end{equation}
Distances between random vectors are then computed by applying the Wasserstein distance to the corresponding distributions in Aitchison coordinates.

\subsection{Wasserstein distance between angular measures}\label{sec:wasserstein_distance}

To compare the extremal dependence structures associated with different random vectors, we compare the empirical measures either directly on the simplex or after transformation to Aitchison coordinates using Wasserstein distances from optimal transport theory. Intuitively, the Wasserstein distance quantifies the minimal cost required to transport one probability distribution onto another. Let $P_i$ and $P_l$ denote two probability measures of random vectors $i$ and $l$, respectively. The $1$-Wasserstein distance between $P_i$ and $P_l$ is \citep{villani2009optimal}
\[
W_1(P_i,P_l)
=
\inf_{\pi\in\Pi(P_i,P_l)}
\int_{\Rset^{p-1}\times\Rset^{p-1}}
\left\|x-y\right\|_2
\,{\rm d}\pi(x,y),
\]
where $\Pi(P_i,P_l)$ denotes the collection of all probability measures on $\Rset^{p-1}\times\Rset^{p-1}$ with marginals $P_i$ and $P_l$. The quantity $\left\|x-y\right\|_2$ represents the transportation cost between the points $x$ and $y$.

The Wasserstein distance therefore provides a comparison between the distributions by taking into account the global organization of probability mass. In contrast with the partition-based approach of \citet{vignotto2021clustering}, this methodology does not require the specification of a partition parameter and avoids the information loss induced by discretizing the simplex.

For arbitrary $p\geq 2$, the empirical Wasserstein distance can be evaluated as a finite-dimensional discrete optimal-transport problem. Let
\[
k_i=\lvert\mathcal T_i\rvert,
\qquad
k_l=\lvert\mathcal T_l\rvert,
\]
be the numbers of retained observations. Re-index the observations in the chosen representation as
\[
y_{i,1},\ldots,y_{i,k_i}
\qquad\text{and}\qquad
y_{l,1},\ldots,y_{l,k_l},
\]
where $y_{i,r}$ denotes either an angular observation on the simplex or its Aitchison coordinates. The corresponding empirical measures are
\[
\widehat P_i
=
\frac{1}{k_i}
\sum_{r=1}^{k_i}\delta_{y_{i,r}},
\qquad
\widehat P_l
=
\frac{1}{k_l}
\sum_{s=1}^{k_l}\delta_{y_{l,s}}.
\]
Define the set of admissible transport plans by
\[
\Gamma_{il}
=
\left\{
\gamma\in\Rset_{+}^{k_i\times k_l}:
\sum_{s=1}^{k_l}\gamma_{rs}
=
\frac{1}{k_i},
\quad r=1,\ldots,k_i,
\qquad
\sum_{r=1}^{k_i}\gamma_{rs}
=
\frac{1}{k_l},
\quad s=1,\ldots,k_l
\right\}.
\]
The empirical $1$-Wasserstein distance is then
\begin{equation}
\widehat d_{il}^{W}
=
W_1(\widehat P_i,\widehat P_l)
=
\min_{\gamma\in\Gamma_{il}}
\sum_{r=1}^{k_i}
\sum_{s=1}^{k_l}
\gamma_{rs}
d\left(y_{i,r},y_{l,s}\right),
\label{eq:empirical-wasserstein-discrete}
\end{equation}
where $d$ is the Euclidean distance in the chosen representation and $\gamma_{rs}$ is the amount of probability mass transported from $y_{i,r}$ to $y_{l,s}$. The marginal constraints ensure that $\gamma$ has $\widehat P_i$ and $\widehat P_l$ as its two marginals.
We evaluate Equation~\eqref{eq:empirical-wasserstein-discrete} numerically using the network-flow implementation of the discrete Kantorovich problem provided by the \texttt{transport} package in \texttt{R} \citep{peyre2019computational,schuhmacher2024transport}. Finally, the dissimilarity matrix,
\begin{equation}
\widehat\bD_T^{W}
=
\left(
\widehat d_{il}^{W}
\right)_{i,l=1,\ldots,n}
\in
\Rset^{n\times n},
\label{eq:angular_dissimilarity_matrix}
\end{equation}
serves as input to any distance-based clustering algorithm.

\section{Consistency of Clustering Procedures}\label{sec:Consistency}

We establish the consistency of the proposed Wasserstein clustering methodology for hierarchical clustering with single, complete, or average linkage and for the global $K$-medoids solution in a projected Aitchison-coordinate space. Recall that $\Delta_{p-1}^{\circ}$ is the open unit simplex defined in equation~\eqref{eq:open_unit_simplex}, and that $\mathcal A:\Delta_{p-1}^{\circ}\to\Rset^{p-1}$ is the Aitchison-coordinate map in equation~\eqref{eq:aitchison_coordinates}. We primarily advocate the Aitchison transformation in the bivariate setting, where it substantially improves clustering performance. In higher dimensions, working directly on the simplex can also be effective, making the transformation optional.

The consistency argument for the direct-simplex approach is immediate. Because the simplex has finite Euclidean diameter, weak convergence of the empirical angular measures implies convergence in $W_1$. The clustering-consistency result then follows from the same separation argument used below, without any projection. The Aitchison case requires additional care because $\mathcal A$ maps the bounded simplex onto the unbounded space $\Rset^{p-1}$: angular observations approaching the simplex boundary are mapped arbitrarily far from the origin, so weak convergence alone need not imply convergence in $W_1$ after transformation. To avoid imposing additional moment conditions, we project the Aitchison coordinates onto a closed ball of radius $M$. The projection is only needed for the proof. For any finite dataset, one can always choose $M$ large enough that the projection does not change any observation.

For $M>0$, let
$$
\mathcal B_M=\{\ba\in\Rset^{p-1}:\|\ba\|_2\leq M\}
$$
and define the Euclidean projection $\Pi_M:\Rset^{p-1}\to\mathcal B_M$ by
$$
\Pi_M(\ba)
=
\begin{cases}
\ba, & \|\ba\|_2\leq M,\\[1mm]
M\ba/\|\ba\|_2, & \|\ba\|_2>M.
\end{cases}
$$
The map $\Pi_M$ is continuous and satisfies
$|\Pi_M(\ba)|_2=\min\{|\ba|_2,M\}$.
\begin{assumption}\label{as:limit}
After marginal standardization to a common Pareto scale, each random vector $\bX_i$, $i\in\{1,\ldots,n\}$, admits a limiting angular probability measure $H_i$ on $\Delta_{p-1}^{\circ}$. Let
\[
G_i=\mathcal A_{\#}H_i
\]
be its image in Aitchison coordinates, where $\mathcal A_{\#}H_i$ denotes the pushforward of $H_i$ under $\mathcal A$.
\end{assumption}
\begin{assumption}\label{as:threshold}
The radial threshold used to construct $\widehat H_{i,T}$ satisfies
$$
r_{i,T}=\frac{T}{\ell_{i,T}},
\qquad
\ell_{i,T}\to\infty,
\qquad
\frac{\ell_{i,T}}{T}\to0,
\qquad T\to\infty.
$$
\end{assumption}
% Under Assumptions~\ref{as:limit} and~\ref{as:threshold} and the sampling conditions stated above,
% $$
% \widehat H_{i,T}\longrightarrow H_i
% $$
% weakly in probability for every $i$; see Proposition~3.3 of \citet{janssen2020kmeans}.
\begin{assumption}\label{as:groups}
The latent groups are characterized by their limiting angular measures:
$$
g(i)=g(l)
\quad\Longleftrightarrow\quad
H_i=H_l.
$$
\end{assumption}
% Because $\mathcal A$ is one-to-one on $\Delta_{p-1}^{\circ}$, this is equivalent to
% $$
% g(i)=g(l)
% \quad\Longleftrightarrow\quad
% G_i=G_l.
% $$
\begin{assumption}\label{as:K}
The number of latent groups $K$ is known and fixed.
\end{assumption}

Let $\widehat G_{i,T}$ be the empirical Aitchison-coordinate measure defined in equation~\eqref{eq:empirical_aitchison_measure}. For $M>0$, define
$$
G_i^{(M)}=(\Pi_M)_{\#}G_i,
\qquad
\widehat G_{i,T}^{(M)}=(\Pi_M)_{\#}\widehat G_{i,T}.
$$
These measures are supported on $\mathcal B_M$ and thus have first moments bounded by $M$.

\begin{lemma}[Preservation of group identifiability]
\label{lem:projected_identifiability}
Under Assumption~\ref{as:groups}, there exists $M_0<\infty$ such that, for every $M\geq M_0$ and every $i,l\in{1,\ldots,n}$,
$$
g(i)=g(l)
\quad\Longleftrightarrow\quad
G_i^{(M)}=G_l^{(M)}.
$$
Consequently,
$$
\Delta_M
=
\min_{g(i)\neq g(l)}
W_1\left(G_i^{(M)},G_l^{(M)}\right)
>0.
$$
\end{lemma}
Define
$$
\widehat d_{il,T}^{W,M}
=
W_1\left(
\widehat G_{i,T}^{(M)},
\widehat G_{l,T}^{(M)}
\right),
\qquad
\widehat\bD_T^{W,M}
=
\left(\widehat d_{il,T}^{W,M}\right)_{1\leq i,l\leq n}.
$$

\begin{theorem}[Consistency of projected Wasserstein--Aitchison clustering]
\label{thm:consistency}
Fix $M\geq M_0$ and let the partition $\widehat{\mathcal P}_T^{(M)}$ of $\{1,\ldots,n\}$ be obtained from $\widehat\bD_T^{W,M}$ by either:
\noindent
(i) hierarchical clustering with single, complete, or average linkage, cut to produce $K$ clusters; or
\noindent
(ii) assignment to the nearest medoid in a global solution
$$
\widehat{\mathcal S}_T^{(M)}
\in
\operatorname*{arg\,min}_{
\substack{
\mathcal S\subseteq\{1,\ldots,n\}\\
\lvert\mathcal S\rvert=K
}}
\sum_{i=1}^{n}
\min_{s\in\mathcal S}
\widehat d_{is,T}^{W,M}.
$$
Under Assumptions~\ref{as:limit}--\ref{as:K},
$$
\prob\left\{
\widehat{\mathcal P}_T^{(M)}=\mathcal P
\right\}
\longrightarrow1,
\qquad T\to\infty,
$$
where $\mathcal P$ is the true latent-group partition.
\end{theorem}
The proofs of Lemma~\ref{lem:projected_identifiability} and Theorem~\ref{thm:consistency} are given in Section~\ref{sec:proof_consistency} of the Supplementary Material.

\section{Simulation Study}\label{sec:Simulation}

The simulation study is designed to evaluate the clustering performance of the proposed Wasserstein--Aitchison methodology and to compare it with the approaches of \citet{vignotto2021clustering} and \citet{o2025clustering} in a wide range of controlled scenarios of extremal dependence. Details on the two competitors can be found in Section~\ref{sec:Competitors} of the Supplementary Material. Clustering agreement is measured using the adjusted Rand index; details on this metric are provided in Section~
\ref{sec:supp_simulation_ARI} of the Supplementary Material.

Because the simulation study involves only a finite collection of finite samples, a common finite projection radius $M$ can be chosen larger than the maximum norm of all observed Aitchison coordinates. The projection then leaves every coordinate unchanged, so the projected and unprojected Wasserstein--Aitchison procedures coincide exactly for all results reported below.

In the simulation studies and applications, the $K$-medoids partition is computed using partitioning around medoids \citep{kaufman1990finding}, an iterative algorithm for approximating the global $K$-medoids solution considered in Theorem~\ref{thm:consistency}. Partitioning around medoids is used because it produced better empirical results than hierarchical clustering in our experiments and was also employed by \citet{vignotto2021clustering} and \citet{o2025clustering}.

The methodology of Section~\ref{sec:Methods} and the consistency result of Section~\ref{sec:Consistency} are stated for a general dimension $p\geq 2$. Simulation studies I and II below are bivariate, whereas simulation study III is trivariate. Mean adjusted Rand indices, Monte Carlo standard errors and exact-recovery rates for all simulation are reported in Tables~\ref{tab:simulation1_mean_ari}--~\ref{tab:simulation3_exact_recovery} of the Supplementary Material.

\subsection{Simulation study I: varying strength of extremal dependence}
For every simulation setting, the latent number of groups is fixed to $K=3$. The study varies the separation between dependence regimes, the sample size, the cluster-size configuration and the copula family. Extremal dependence is summarized by the upper tail dependence coefficient $\chi$, defined for $p=2$ by
\[
\chi = 2\int_0^1 \min(\theta,1-\theta)\,dH(\theta).
\]
Three levels of separation between the latent groups are considered: easy, with $\chi=(0.1,0.5,0.9)$, yielding clearly separated extremal dependence structures; moderate, with $\chi=(0.2,0.5,0.8)$, producing a more challenging clustering problem; and hard, with $\chi=(0.3,0.45,0.6)$, corresponding to weakly separated extremal dependence structures.

We consider four different sample sizes, namely $T\in\{250,500,750,\num{1000}\}$. Two cluster-size configurations are considered: a balanced design with $N=(10,10,10)$, in which each latent group contains the same number of random vectors, and an unbalanced design with $N=(5,10,20)$. The simulations are conducted under two bivariate extreme-value copula families, the Gumbel--Hougaard and the H\"usler--Reiss copula. The parameter $W$ which controls the partitioning of the space, is set to $W\in\{2,3,4,5,8,10,15\}$. This allows us to assess the sensitivity of the method to the partition resolution. For both methods, extremal dependence is estimated using only the largest $10\%$ of radial observations and the conditional extremes approach is fitted above the $90$th percentile threshold. Finally, for each scenario, $500$ independent replications are generated.
\begin{figure}
    \centering
    \includegraphics[
        width=\textwidth,
        height=0.7\textheight,
        keepaspectratio
    ]{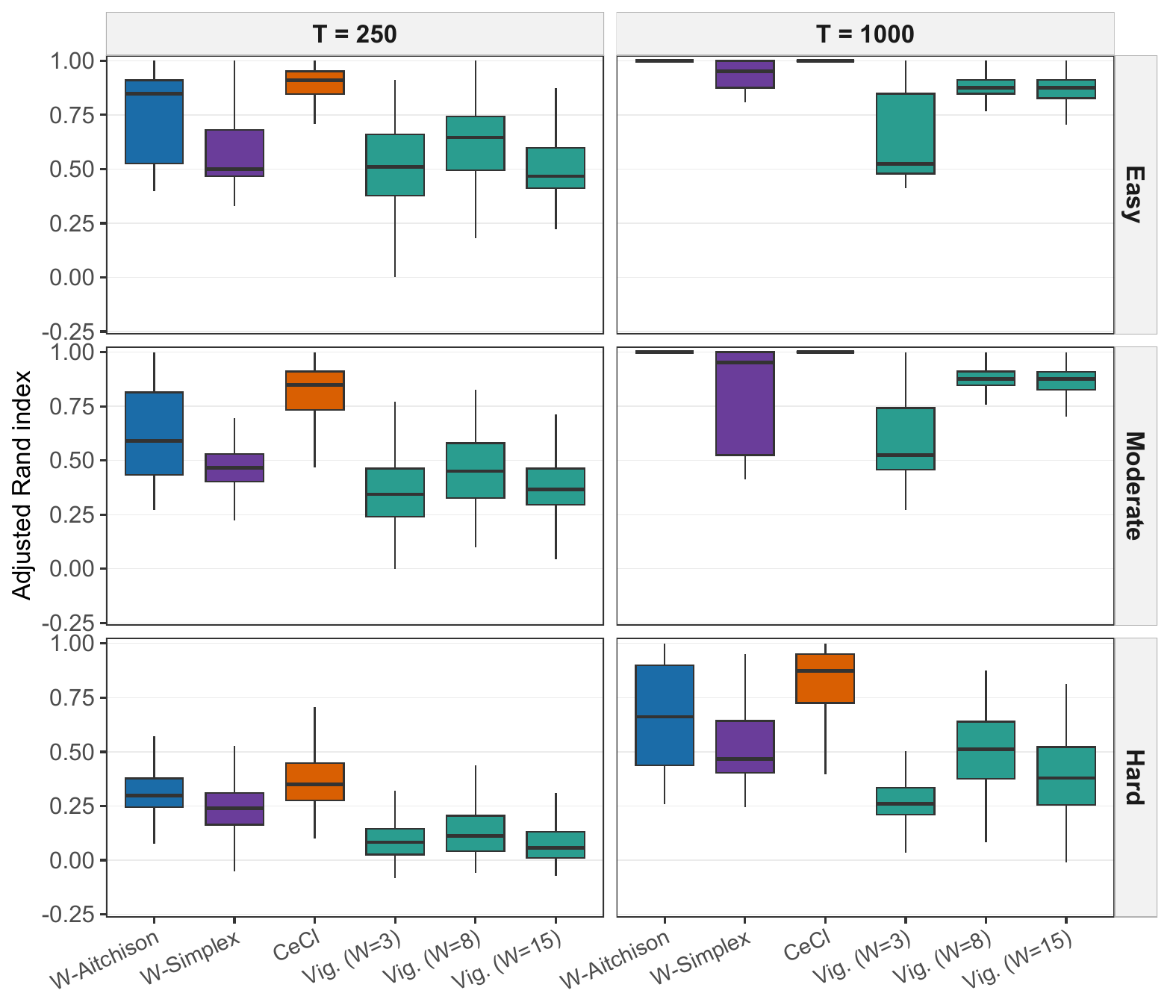}
    \caption{Adjusted Rand index for a subset of the unbalanced design $N=(5,10,20)$, with $K=3$ latent groups, for sample sizes $T\in\{250,\num{1000}\}$ and the three separation levels easy, moderate and hard, corresponding to $\chi=(0.1,0.5,0.9)$, $(0.2,0.5,0.8)$ and $(0.3,0.45,0.6)$, under the H\"usler--Reiss copulas. The Wasserstein--Aitchison, Wasserstein--simplex and conditional extremes dissimilarities are compared with the partition-based dissimilarity of \citet{vignotto2021clustering}, the latter for $W\in\{3,8,15\}$ partitions. Results are based on 500 Monte Carlo replications.}
    \label{fig:ari_unbalanced_main}
\end{figure}

Figure~\ref{fig:ari_unbalanced_main} report the adjusted Rand indices for a subset of the unbalanced design. While, Figures~\ref{fig:ari_unbalanced} and~\ref{fig:ari_balanced} in Section~\ref{sec:supp_simulation} of the Supplementary Material report it for the full design. We observed that Wasserstein--Aitchison and conditional extremes clustering consistently outperform the partition-based method of \citet{vignotto2021clustering}. Although the latter improves as the number of partitions increases, particularly from small values, it rarely attains the accuracy of the other two methods. Performance also improves substantially with sample size: in the easy and moderate scenarios, both leading methods approach perfect recovery for the Gumbel and H\"usler--Reiss models under both designs. Their performance is similar for balanced groups, whereas conditional extremes clustering is generally more accurate under unbalanced designs, especially in the moderate and hard scenarios. As expected, decreasing separation from the easy to the hard setting reduces accuracy and increases variability. Finally, Wasserstein--Aitchison outperforms Wasserstein--simplex, indicating that the Aitchison transformation provides a more informative geometry for distinguishing bivariate extremal dependence structures.

\subsection{Simulation study II: equal tail dependence and different angular shapes}

The first simulation study considers groups with different upper tail-dependence coefficients $\chi$. Because $\chi$ does not fully characterize the angular measure, we now examine whether the procedures can distinguish groups having the same $\chi$ but different angular structures. The three groups are generated from Galambos, H\"usler--Reiss and extremal-$t$ copulas, the latter with $\nu=5$ degrees of freedom. For each $\chi\in\{0.1,0.3,0.5,0.7,0.9\}$, the copula parameters are calibrated to give the same value of $\chi$ across groups. Their recovery must therefore rely on differences between the complete angular measures rather than on the overall strength of tail dependence. Given the difficulty of this setting, we consider $T\in\{\num{10000},\num{100000}\}$.

We compare Wasserstein--Aitchison and conditional extremes clustering, both combined with $K$-medoids and the true value $K=3$. The dissimilarity of \citet{vignotto2021clustering} is omitted because it was dominated by both methods in the first simulation study. Wasserstein--Aitchison retains the largest $10\%$ of radial observations, while the conditional extremes model is fitted above the $90$th percentile. Each scenario is evaluated using 500 independent Monte Carlo replications.

\begin{figure}
    \centering
    \includegraphics[width=\textwidth]{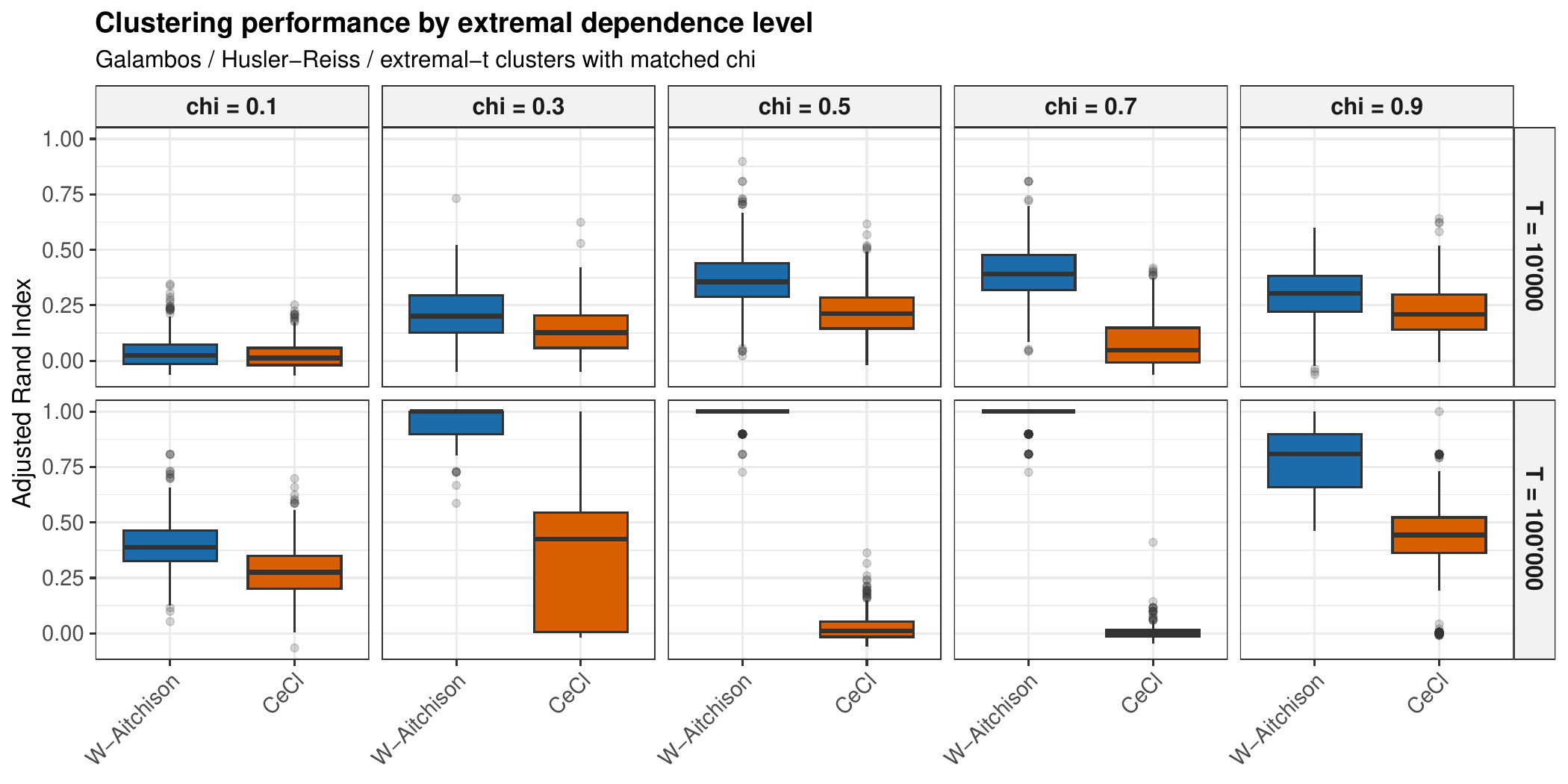}
    \caption{Adjusted Rand index obtained by the Wasserstein--Aitchison and conditional extremes clustering procedures for the equal-tail-dependence simulation study. The three latent groups are generated from Galambos, H{\"u}sler--Reiss and extremal-$t$ copulas calibrated to share the same upper tail dependence coefficient $\chi$. Results are shown for $\chi\in\{0.1,0.3,0.5,0.7,0.9\}$ and sample sizes $T\in\{\num{10000},\num{100000}\}$ based on 500 Monte Carlo replications.}
    \label{fig:ARI_sim2}
\end{figure}

Figure~\ref{fig:ARI_sim2} reports the adjusted Rand index obtained by the two procedures. The Wasserstein--Aitchison method outperforms conditional extremes clustering in this setting as well, attaining a higher adjusted Rand index. The difference between the two procedures is most pronounced for the larger sample size $T=\num{100000}$ and for intermediate levels of extremal dependence, that is, for $\chi$ around $0.5$. Since all three latent groups share the same upper tail dependence coefficient by construction, this difference cannot be attributed to the strength of extremal dependence. It indicates instead that comparing the complete angular measures, rather than a fitted conditional model, retains more of the information contained in the shape of the angular measure. 

\subsection{Simulation study III: trivariate asymmetric angular measures}
\label{sec:simulation_3}
The third simulation study examines whether the Wasserstein--Aitchison and Wasserstein--simplex procedures remain effective in dimension $p=3$, where angular measures are supported on the two-dimensional simplex embedded in $\Rset^3$. We consider three asymmetric angular measures that have identical pairwise tail-dependence coefficients, $\chi_{12}=\chi_{13}=\chi_{23}=0.45$,
but differ in their orientation and higher-order dependence structure. Consequently, the latent groups cannot be distinguished using pairwise tail-dependence coefficients alone.

\begin{figure}
    \centering
    \includegraphics[
        width=0.8\textwidth
    ]{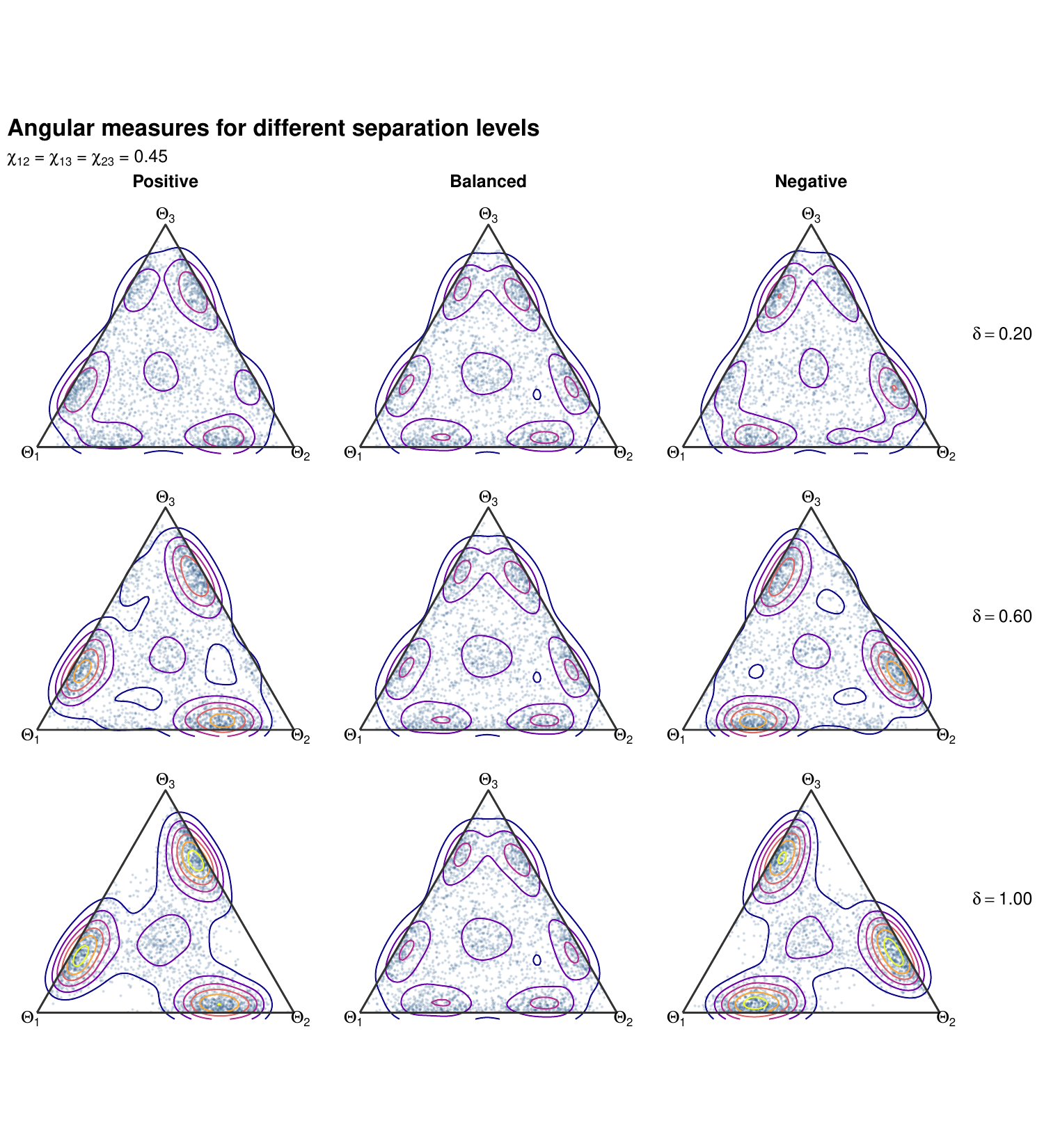}
    \caption{Angular measures used in the trivariate simulation study. Rows correspond to the separation parameter $\delta\in\{0.20,0.40,0.60,0.80,1\}$. Columns correspond, respectively, to the positive-orientation, orientation-balanced and negative-orientation angular measures. Points represent independent draws from the angular measures, while contours indicate regions of greater angular density. All panels have identical pairwise tail-dependence coefficients, $\chi_{12}=\chi_{13}=\chi_{23}=0.45$.}
    \label{fig:simplex_delta_orientation}
\end{figure}

The three angular measures are denoted by $H_{+,\delta}$, $H_0$ and $H_{-,\delta}$, where $\delta\in\{0.20,0.60,1\}$ controls their separation. For small $\delta$, the positive- and negative-orientation measures are close to $H_0$, whereas $\delta=1$ yields their greatest separation, without changing their pairwise tail-dependence coefficients. The measures are constructed using constrained mixtures of Dirichlet densities; full details are provided in Section~\ref{sec:trivariate_construction} of the Supplementary Material. Observations are generated by drawing an angle from the relevant measure and multiplying it by an independent Pareto radial component. Figure~\ref{fig:simplex_delta_orientation} displays \num{3000} independent angular observations from each measure for every value of $\delta$.

The three vertices of the simplex represent extreme events dominated by a single component. For example, observations close to the first vertex satisfy $\Theta_1\approx 1$ and $\Theta_2,\Theta_3\approx 0$. Observations close to the centre of the simplex correspond to events in which all three components contribute similar proportions to the overall magnitude.

The middle column does not change across rows because it always represents $H_0$. In the left and right columns, the angular measures move progressively farther from $H_0$ as $\delta$ increases, and they do so in opposite directions: their orientations, in the sense made precise in Section~
\ref{sec:trivariate_construction} of the Supplementary Material, have opposite signs. For small values of $\delta$, the three columns are visually similar because $H_{+,\delta}$ and $H_{-,\delta}$ are both dominated by the common component $H_0$. When $\delta=1$, the differences between the three angular measures are the most pronounced. 

For each value of $\delta$, one simulated dataset contains three latent groups, with four random vectors per group and $T=\num{1000}$ independent observations per random vector. For each random vector, the largest $10\%$ of the observations according to their radial component $\lVert\bZ_{i,t}\rVert_1$ are retained to compute the empirical angular measure.

The experiment is repeated independently 500 times for each of the five values of $\delta$. For each dataset, the proposed Wasserstein--Aitchison and Wasserstein--simplex dissimilarities and the conditional extremes dissimilarity are computed. The resulting dissimilarity matrices are then clustered using partitioning around medoids with the true number of groups, $K=3$. 

\begin{figure}
\centering
\includegraphics[width=0.9\textwidth]{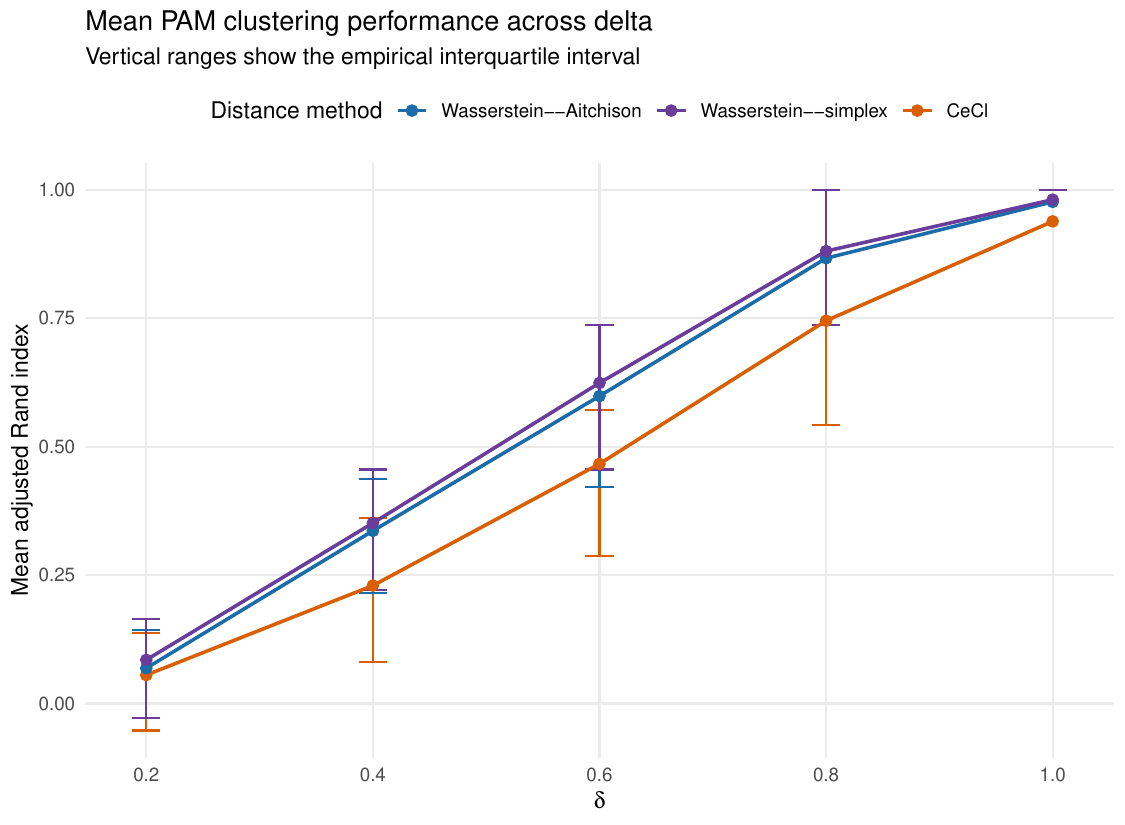}
\caption{Mean adjusted Rand index for partitioning around medoids using the Wasserstein--Aitchison, Wasserstein--simplex and conditional extremes dissimilarities. Results are based on 500 replications; vertical ranges are empirical interquartile intervals. For every $\delta$, the three groups have identical first angular moments and pairwise coefficients $\chi_{12}=\chi_{13}=\chi_{23}=0.45$.}
\label{fig:delta_mean_ari}
\end{figure}

Figure~\ref{fig:delta_mean_ari} compares the three dissimilarities as a function of $\delta$.
Performance improves monotonically with $\delta$. For $\delta=0.20$, the angular measures remain close to $H_0$, and the mean adjusted Rand indices are only $0.069$, $0.085$ and $0.055$ for Wasserstein--Aitchison, Wasserstein--simplex and conditional extremes clustering, respectively. For $\delta=0.40$, $0.60$ and $0.80$, the corresponding values are $(0.336,0.351,0.229)$, $(0.598,0.624,0.466)$ and $(0.867,0.881,0.745)$. Thus, both Wasserstein methods outperform conditional extremes clustering at intermediate separation levels. Unlike in the bivariate experiments, Wasserstein--simplex performs similarly to Wasserstein--Aitchison: their mean adjusted Rand indices differ by at most $0.026$.

At $\delta=1$, the mean adjusted Rand indices reach $0.977$, $0.981$ and $0.939$, with perfect recovery in $91.4\%$, $93.0\%$ and $78.8\%$ of the replications. Overall, the two Wasserstein formulations perform nearly equivalently and consistently outperform conditional extremes clustering. This advantage arises from comparing the complete trivariate angular measures.

\section{Data Application}\label{sec:Application}

\subsection{Application to international equity markets}

We now illustrate the proposed Wasserstein--Aitchison clustering procedure on international equity-market data. The objective is to cluster national stock-market indices according to their extremal co-movement with the United States market. For each country $i$, we consider the bivariate vector $(-R_{\mathrm{US},t},-R_{i,t})$, where $R_{\mathrm{US},t}$ denotes the daily log-return of the S\&P 500 index and $R_{i,t}$ denotes the daily log-return of the equity index associated with country $i$. Positive values therefore correspond to market losses. The analysis focuses on the extremal dependence between large United States market losses and large country-specific losses. Daily closing prices are downloaded from Yahoo Finance for a broad set of national equity indices covering North America, South America, Europe, Asia--Pacific, the Middle East and Africa. The sample period extends from January 2000 until June 2025. Only indices with at least \num{1000} valid daily observations are retained. Daily log-returns are then computed from the cleaned price series.

Since financial returns typically exhibit serial dependence and volatility clustering, the return series are filtered prior to the extremal analysis. For each country, both the United States and local-market returns are fitted separately using a first-order autoregressive model with generalized autoregressive conditional heteroscedasticity of order $(1,1)$ and Student $t$ innovations. The standardized residuals are then extracted and used as inputs for the clustering procedure. This filtering step aims to remove temporal dependence and conditional heteroscedasticity.

For each country, a bivariate sample is constructed by pairing the filtered United States residuals with the filtered residuals of the corresponding local market, $\left(\widehat{\varepsilon}_{\mathrm{US},t},\widehat{\varepsilon}_{i,t}\right)$. The methodology described in the previous sections is then applied. Observations are transformed to Pareto margins, radial--angular coordinates are computed, and only the largest $5\%$ of radial observations are retained. The resulting angular observations are transformed into Aitchison coordinates and pairwise Wasserstein distances are computed between countries. The resulting dissimilarity matrix is finally clustered using partitioning around medoids.

The number of clusters is selected using the average silhouette width. Figure~\ref{fig:silhouette_selection_finance} in Section~
\ref{sec:supplementary_finance} of the Supplementary Material displays the silhouette criterion for values of $K$ ranging from 2 to 10. The maximum silhouette value is obtained at $K=2$, indicating that the dominant structure in the data is a two-cluster partition.

\begin{figure}
    \centering
    \includegraphics[width=0.85\textwidth]{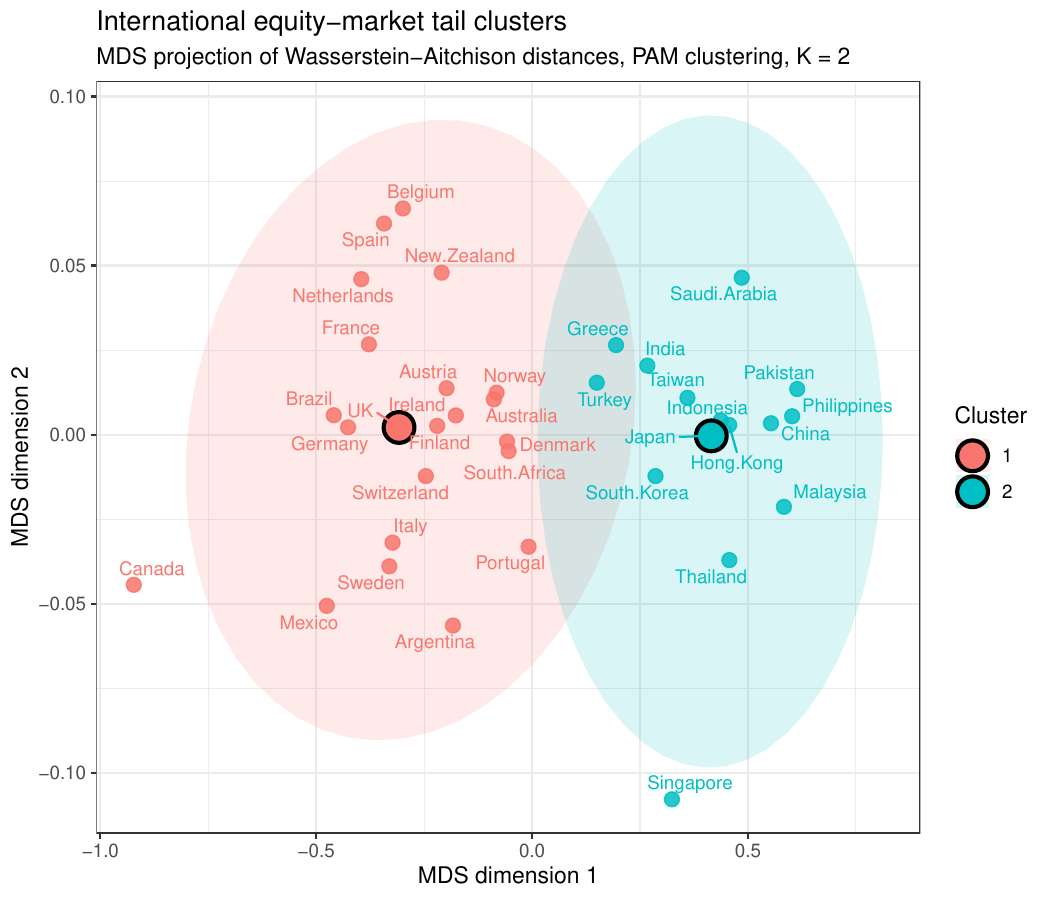}
    \caption{International equity-market tail clusters obtained from the Wasserstein--Aitchison dissimilarity matrix. The figure shows a two-dimensional classical multidimensional scaling projection of the pairwise distances together with the partitioning around medoids result. Black-circled observations correspond to the cluster medoids.}
    \label{fig:finance_clusters}
\end{figure}

Figure~\ref{fig:finance_clusters} presents a two-dimensional classical multidimensional scaling representation of the Wasserstein--Aitchison distance matrix together with the partitioning around medoids results. The two clusters exhibit a clear geographical and economic interpretation. The first cluster contains most North American, South American and European markets together with Australia and New Zealand. The second cluster is composed primarily of Asian markets, including China, Hong Kong, India, South Korea, Taiwan, Singapore, Malaysia, Thailand, Indonesia, the Philippines and Pakistan. The cluster medoids are the United Kingdom and Japan.

The two medoids show distinct extremal-dependence regimes. The angular distribution of the United Kingdom is more concentrated near the centre of the simplex, indicating a relatively balanced sharing of extreme losses between the United States and United Kingdom markets, whereas that of Japan has greater mass near the boundaries, indicating that extreme events are more frequently dominated by one market. Detailed angular-density estimates for the medoids and for countries near the cluster boundaries are provided in Figs.~\ref{fig:finance_medoids} and \ref{fig:finance_boundary_countries} in Section~\ref{sec:supplementary_applications} of the Supplementary Material. Applying the conditional extremes clustering methodology of \citet{o2025clustering} to the same data set yields exactly the same two-cluster partition; the corresponding multidimensional scaling representation is shown in Fig.~\ref{fig:cecl_country_clusters} of the Supplementary Material. This agreement supports the robustness of the identified extremal-dependence structure.

\subsection{Clustering rain--snowmelt--runoff extremal dependence in hydrologic Switzerland}
\label{sec:Trivariate application}

This application investigates whether Swiss catchments can be grouped according to the extremal dependence structure linking liquid precipitation, snowpack depletion and runoff response. We use version 0.8 of the Catchment Attributes and Meteorology for Large-sample Studies--Switzerland data set (CAMELS-CH) \citep{hoge2023camelsch,hoge2023camelschdata}. For catchment $i$ and day $t$, the analysis is based on the trivariate vector
\[
X_{i,t}
=
\left(
L_{i,t},
M_{i,t},
Q^{\star}_{i,t}
\right),
\]
whose components represent three aspects of a potential flood-generating event. First, $L_{i,t}$ is liquid precipitation accumulated over the current and two preceding days, obtained by partitioning daily precipitation into liquid and solid fractions using air temperature. Second, $M_{i,t}$ is the positive net depletion of snow water equivalent over three days and is used as a proxy for snowpack water release. Third, $Q^{\star}_{i,t}$ is the maximum observed specific discharge on day $t$ or either of the following two days, allowing for a delay between meteorological forcing and runoff response. Three day accumulation was chosen based on the typical passage time of extratropical cyclones \citep{froidevaux2015flood}. The analysis covers 1 September 1998--31 December 2020, determined by the availability of snow-water-equivalent data. Complete definitions and data-quality criteria are provided in Section~\ref{sec:supplementary_swiss} of the Supplementary Material.

Within each catchment, temporal dependence among extreme observations was handled using runs declustering with a run length of three days, following \citet{sabourin2015combining}. An event was considered to have ended after three consecutive days without a radial exceedance, and the observation with the largest radial component within each event was retained. We retained only catchments with at least 50 events, a deliberately low minimum for estimating a trivariate angular measure. We applied the proposed Wasserstein--Aitchison methodology to the upper $5\%$ of each catchment-specific radial distribution. The average silhouette width slightly favoured $K=3$ over $K=2$ (Fig.~\ref{fig:camels_sensitivity_silhouette}), but the sensitivity analysis in Section~\ref{sec:supplementary_swiss} of the Supplementary Material showed substantially greater stability for $K=2$, which we therefore retained.

\begin{figure}
    \centering
    \includegraphics[width=\textwidth]
    {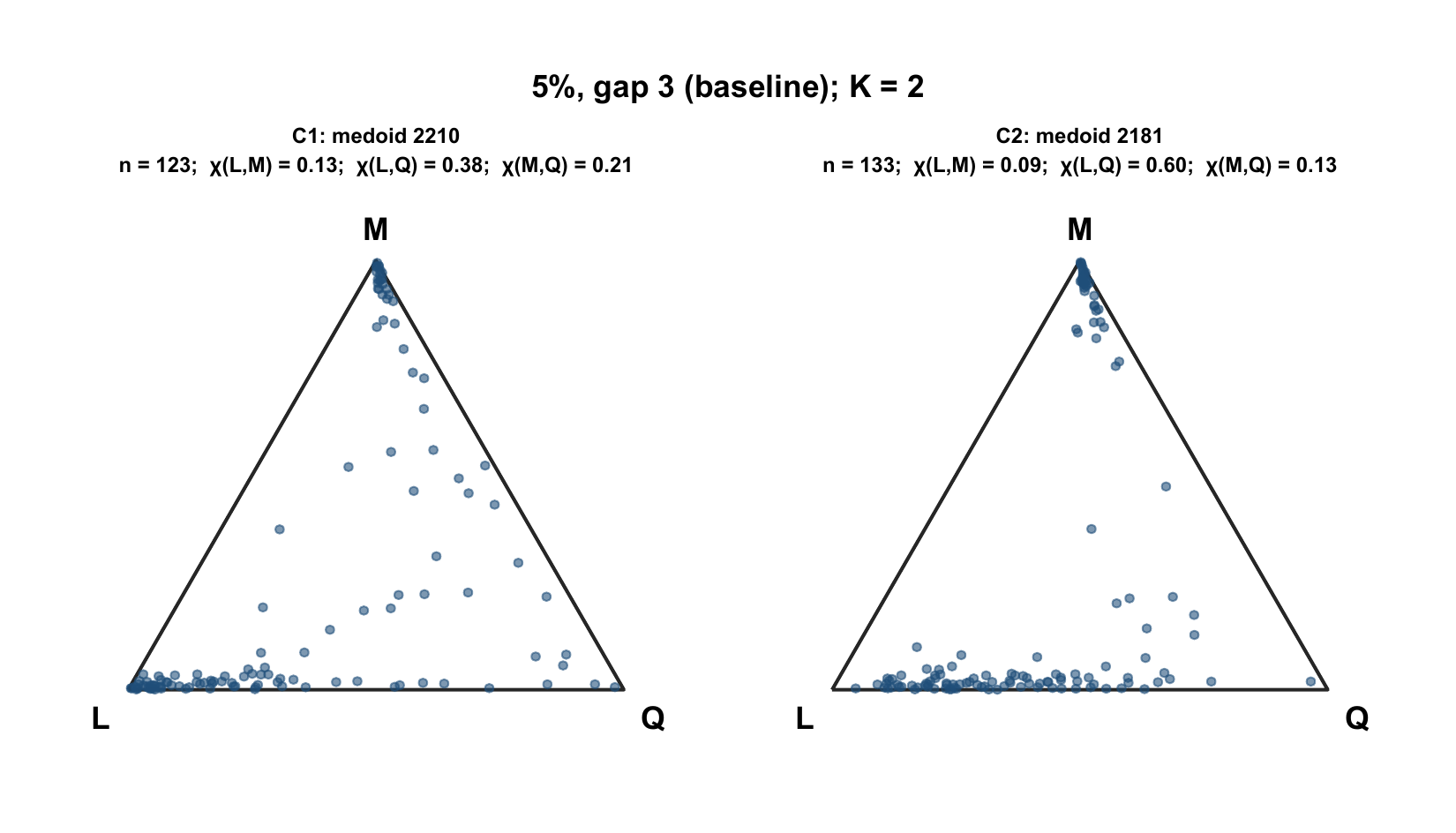}
    \caption{Empirical angular measures of the $K=2$ cluster medoids (C1, left; C2, right). Here $L$, $M$ and $Q$ denote standardized liquid precipitation, snow-water-equivalent depletion and runoff, respectively. Annotations report pairwise tail-dependence coefficients and numbers of declustered events.}
    \label{fig:camelsch_simplex}
\end{figure}

The resulting partition exhibits broad spatial organization without forming strictly contiguous regions, as shown in Fig.~\ref{fig:camelsch_map} of the Supplementary Material. Because geographical coordinates and catchment adjacency were not used in the clustering, this organization indicates an association with regional hydroclimatic conditions rather than a direct reproduction of geographical proximity. Figure~\ref{fig:camelsch_simplex} displays the empirical angular measures of the two cluster medoids and reveals two distinct extremal-dependence regimes. The C1 medoid has the most broadly dispersed angular measure, with substantial mass in rainfall-related directions but also comparatively frequent snowmelt-related events. In contrast, the C2 medoid is strongly concentrated near the liquid-rain--runoff edge and has a substantially larger rain--runoff tail-dependence coefficient than its rain--melt and melt--runoff coefficients. It represents the clearest rainfall--runoff regime.

\begin{figure}
    \centering
    \includegraphics[width=\textwidth]
    {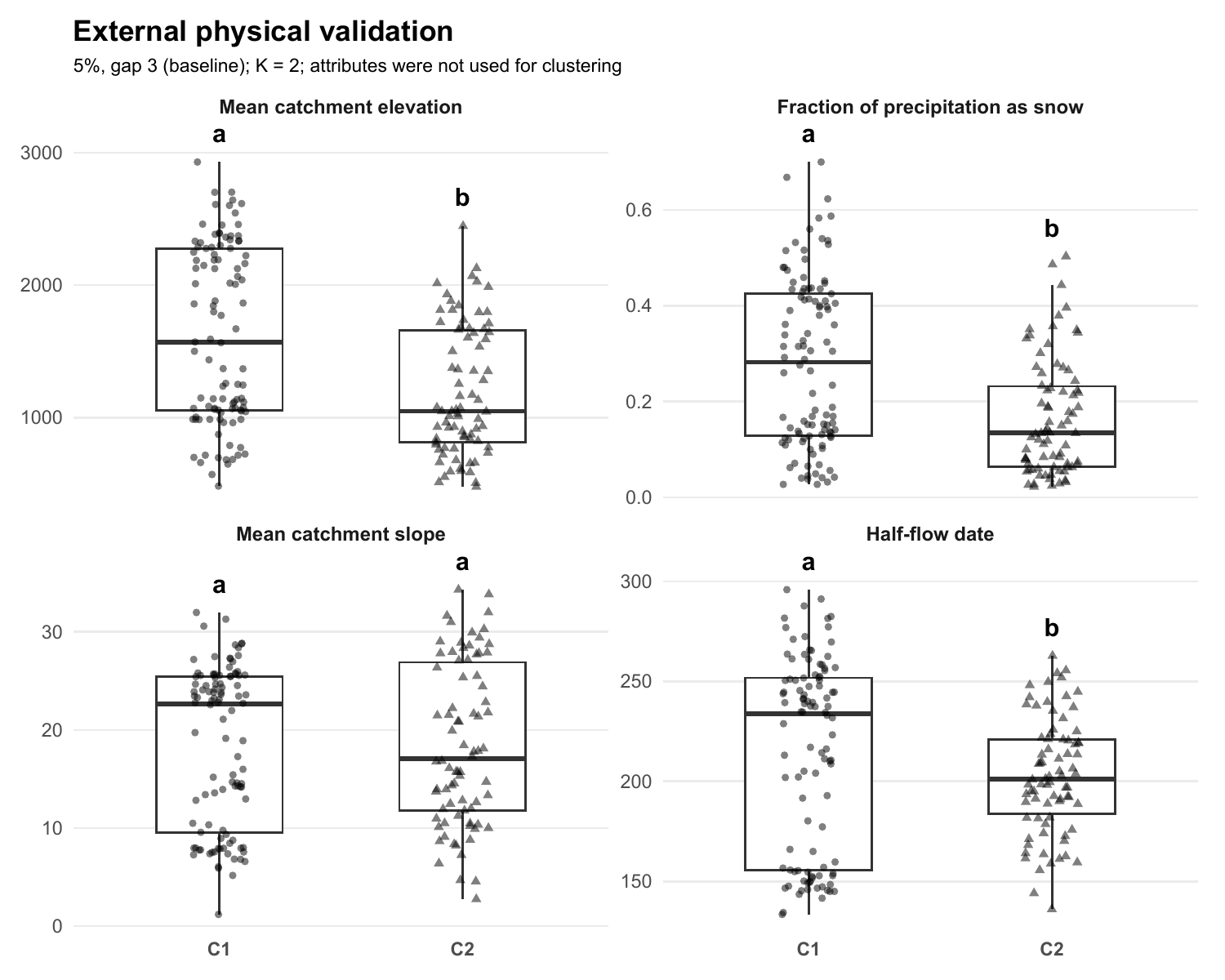}
    \caption{External physical validation of the two extremal-dependence clusters. The panels show mean catchment elevation, fraction of precipitation falling as snow, mean catchment slope and mean half-flow date. Individual symbols represent catchments, with circles, triangles corresponding to C1, C2. Lowercase letters denote significance groups based on pairwise Wilcoxon rank-sum tests at the 5\% level.}
    \label{fig:camelsch_validation}
\end{figure}

The hydrological interpretation is supported by four catchment attributes that were excluded from the clustering: mean elevation, fraction of precipitation falling as snow, mean slope and mean half-flow date. As shown in Fig.~\ref{fig:camelsch_validation}, C1 contains higher and more snow-influenced catchments with later runoff timing; C2 contains lower catchments with smaller snow fractions.

\section{Conclusion}\label{sec:Conclusion}

We have introduced a dissimilarity between multivariate extremal dependence structures based on optimal transport. The angular measure of each random vector is estimated nonparametrically above a high radial threshold, mapped from the unit simplex to $\Rset^{p-1}$ through the centred log-ratio transformation of Aitchison geometry, and two random vectors are compared through the $1$-Wasserstein distance between the resulting empirical measures. The construction requires neither an ad hoc partition of the simplex, as in \citet{vignotto2021clustering}, nor the fitting of a conditional extremes model, as in \citet{o2025clustering}. Theorem~\ref{thm:consistency} shows that hierarchical clustering and global K-medoid based on the resulting dissimilarity matrix recovers the latent partition with probability tending to one. The simulation study of Section~\ref{sec:Simulation} confirms this behaviour in finite samples and shows that the gain over existing procedures is largest precisely in the situations that motivate the approach, namely when the latent groups cannot be separated by their pairwise tail dependence coefficients alone. The two applications of Section~\ref{sec:Application} yield interpretable partitions: a separation of Swiss catchments into more snowmelt-influenced and predominantly rainfall--runoff regimes, and a separation between Western developed markets and Asian and emerging markets according to their extremal co-movement with the US market.

Several directions deserve further attention. First, the radial threshold is chosen as a fixed proportion of the observations, and a data-driven choice, possibly differing across random vectors, would make the procedure more automatic. Second, the consistency result assumes that the number of latent groups $K$ is known, whereas in the applications it is selected by maximizing the average silhouette width; a formal treatment of this selection step is left for future work. Finally, the observations are assumed to be independent over time. In the applications, temporal dependence is removed by prefiltering the series or run declustering, but a direct treatment of serially dependent extremes would be a natural extension.

Possible future applications in climate science could include analysis of heatwave-drought compounding mechanisms and their changes under global warming.

% =============================================================================
% Back matter
% =============================================================================

\backmatter

\bmhead{Supplementary material}

Supplementary material containing additional methodological details,
proofs, simulation results and application results is provided after the
main manuscript.

\section*{Declarations}

\bmhead{Funding}

This research is supported by the Projet d'Actions de Recherche Concertées (ARC) No.~24/29-146 (EXALT) of the Communauté française de Belgique.

\bmhead{Use of generative AI and AI-assisted technologies}

During the preparation of this manuscript, the authors used ChatGPT for coding and debugging assistance, support in checking mathematical proofs, and concise language editing. All AI-assisted output was critically reviewed and verified by the authors, who take full responsibility for the manuscript.
% =============================================================================
% Supplementary material
% =============================================================================
\clearpage
% ============================================================
% Supplementary Material
% ============================================================

\clearpage

\section*{Supplementary Material}

\setcounter{section}{0}
\setcounter{subsection}{0}
\setcounter{figure}{0}
\setcounter{table}{0}
\setcounter{equation}{0}

\renewcommand{\thesection}{S\arabic{section}}
\renewcommand{\thesubsection}{\thesection.\arabic{subsection}}
\renewcommand{\thefigure}{S\arabic{figure}}
\renewcommand{\thetable}{S\arabic{table}}
\renewcommand{\theequation}{S\arabic{equation}}
% Unique hyperlink anchors for the S-numbered objects (hyperref is loaded by sn-jnl)
\renewcommand{\theHsection}{S\arabic{section}}
\renewcommand{\theHsubsection}{\theHsection.\arabic{subsection}}
\renewcommand{\theHfigure}{S\arabic{figure}}
\renewcommand{\theHtable}{S\arabic{table}}
\renewcommand{\theHequation}{S\arabic{equation}}

\section{Extreme-value copulas}\label{sec:ev_copulas}

Since the objective is to model dependence between extreme realizations of the innovation process, we restrict attention to the class of extreme-value copulas. Extreme-value copulas arise naturally as limiting dependence structures of componentwise maxima and constitute a fundamental class of models in multivariate extreme-value theory \citep{gudendorf2010extreme}. They are particularly well suited for describing extremal dependence between variables. In the multivariate setting, every $p$-variate extreme-value copula admits the representation
\[
C(u_1,\dots,u_p)
=
\exp\!\left\{
-\ell\!\left(
-\log u_1,\dots,-\log u_p
\right)
\right\},
\qquad
(u_1,\dots,u_p)\in(0,1]^p,
\]
where $\ell:[0,\infty)^p\to[0,\infty)$ is the stable tail dependence function. The function $\ell$ is convex, homogeneous of order one, and satisfies
\[
\max(x_1,\dots,x_p)
\le
\ell(x_1,\dots,x_p)
\le
x_1+\dots+x_p,
\qquad
(x_1,\dots,x_p)\in[0,\infty)^p.
\]
% By homogeneity, $\ell$ is completely characterized by its restriction to the unit simplex
% \[
% \Delta_{p-1}
% =
% \Bigl\{
% (w_1,\dots,w_p)\in[0,1]^p:
% \sum_{j=1}^p w_j = 1
% \Bigr\},
% \]
% through the Pickands dependence function $A:\Delta_{p-1}\to[1/p,1]$, defined by
% \[
% \ell(x_1,\dots,x_p)
% =
% \left(
% \sum_{j=1}^p x_j
% \right)
% A(w_1,\dots,w_p),
% \qquad
% w_j
% =
% \frac{x_j}{\sum_{k=1}^p x_k}.
% \]
% Consequently,
% \[
% C(u_1,\dots,u_p)
% =
% \exp\!\left\{
% \left(
% \sum_{j=1}^p \log u_j
% \right)
% A\!\left(
% \frac{\log u_1}{\sum_{k=1}^p \log u_k},
% \dots,
% \frac{\log u_p}{\sum_{k=1}^p \log u_k}
% \right)
% \right\}.
% \]
% The Pickands dependence function $A$ is convex and satisfies
% \[
% \max(w_1,\dots,w_p)
% \le
% A(w_1,\dots,w_p)
% \le
% 1,
% \qquad
% (w_1,\dots,w_p)\in\Delta_{p-1}.
% \]
% The upper bound $A(w)=1$ corresponds to independence, whereas the lower bound
% \[
% A(w)=\max(w_1,\dots,w_p)
% \]
% corresponds to complete dependence.

In the bivariate setting, several summary measures can be used to quantify the dependence structure induced by a copula. For a bivariate copula $C$, we consider the upper tail dependence coefficient
\begin{equation}
\chi
=
\lim_{u\uparrow 1}
\operatorname{pr}\!\left(
U_1>u
\mid
U_2>u
\right),
\label{eq:chi}
\end{equation}
provided the limit exists. Equivalently,
\[
\chi
=
\lim_{u\uparrow 1}
\frac{
1-2u+C(u,u)
}{
1-u
}.
\]
The coefficient $\chi$ measures the tendency of the variables to become simultaneously extreme. Its values satisfy $0 \le \chi \le 1$.
The case $\chi = 0$ corresponds to asymptotic independence, whereas $\chi > 0$ indicates asymptotic dependence \citep{joe2014dependence}. For bivariate extreme-value copulas with Pickands dependence function $A$, the coefficient is given by $\chi=2\bigl(1-A(1/2)\bigr)$ \citep{gudendorf2010extreme}.

% 
% , Kendall's tau is defined as
% \begin{equation}
% \tau
% =
% 4
% \int_0^1
% \int_0^1
% C(u_1,u_2)\,
% \mathrm{d}C(u_1,u_2)
% -1.
% \label{eq:kendall_tau}
% \end{equation}
% Kendall's tau measures the overall strength of dependence and takes values in the interval $[-1,1]$ \citep{nelsen2006introduction}.

% Spearman's rho is defined by
% \begin{equation}
% \rho_S
% =
% 12
% \int_0^1
% \int_0^1
% C(u_1,u_2)\,
% \mathrm{d}u_1\,\mathrm{d}u_2
% -3.
% \label{eq:spearman_rho}
% \end{equation}
% Like Kendall's tau, Spearman's rho measures global dependence and satisfies $-1 \le \rho_S \le 1$. Both coefficients are invariant under strictly increasing transformations of the margins and therefore depend only on the copula \citep{nelsen2006introduction}.

% While Kendall's tau and Spearman's rho summarize overall dependence, they do not specifically quantify extremal dependence.

We now present the parametric families of copulas that will be used in the simulation study: the Gumbel--Hougaard, H\"usler--Reiss and Galambos copulas, together with the $t$-copula. The first three belong to the class of extreme-value copulas but exhibit different forms of extremal dependence, whereas the $t$-copula does not.

The Gumbel--Hougaard copula \citep{gumbel1960bivariate,gumbel1961bivariate}, also known as the logistic extreme-value copula, is defined in dimension $p$ by
\begin{equation}
C_\phi(u_1,\dots,u_p)
=
\exp\!\left\{
-
\left(
\sum_{j=1}^{p}
(-\log u_j)^\phi
\right)^{1/\phi}
\right\},
\qquad
\phi \ge 1.
\label{eq:gumbel_copula}
\end{equation}
Its associated stable tail dependence function is
\[
\ell_\phi(x_1,\dots,x_p)
=
\left(
\sum_{j=1}^{p}
x_j^\phi
\right)^{1/\phi},
\]
and the corresponding Pickands dependence function is
\[
A_\phi(w_1,\dots,w_p)
=
\left(
\sum_{j=1}^{p}
w_j^\phi
\right)^{1/\phi},
\qquad
(w_1,\dots,w_p)\in\Delta_{p-1}.
\]
The parameter $\phi$ controls the strength of extremal dependence. The case $\phi=1$ corresponds to independence, whereas $\phi\to\infty$ yields complete dependence. In the bivariate case, the upper tail dependence coefficient admits the closed-form expression
\begin{equation}
\chi
=
2-2^{1/\phi}.
\label{eq:chi_gumbel}
\end{equation}
Consequently, larger values of $\phi$ correspond to stronger extremal dependence.

Another important class of extreme-value copulas is provided by the H\"usler--Reiss family \citep{husler1989maxima}. In dimension $p$, the H\"usler--Reiss copula is parameterized by a variogram matrix
\[
\Gamma=(\Gamma_{ij})_{1\le i,j\le p},
\]
where $\Gamma_{ii}=0$, $\Gamma_{ij}=\Gamma_{ji}\ge0$, and $\Gamma$ is conditionally negative definite. The corresponding stable tail dependence function admits the representation
\[
\ell(x_1,\dots,x_p)
=
\sum_{j=1}^{p}
x_j
\Phi_{p-1}\!\left(
\eta^{(j)};
\Sigma^{(j)}
\right),
\]
where $\Phi_{p-1}$ denotes the $(p-1)$-variate Gaussian distribution function and the vectors $\eta^{(j)}$ and covariance matrices $\Sigma^{(j)}$ are determined by the variogram matrix $\Gamma$ \citep{engelke2020graphical}. The variogram controls the strength of extremal dependence between the components. Small values of $\Gamma_{ij}$ correspond to strong extremal dependence, whereas large values of $\Gamma_{ij}$ correspond to weak extremal dependence. In the bivariate case, the model reduces to a single variogram parameter $\Gamma_{12}=\Gamma$ and the Pickands dependence function becomes
\[
A_\Gamma(t)
=
(1-t)\,
\Phi\!\left(
\frac{\sqrt{\Gamma}}{2}
+
\frac{1}{\sqrt{\Gamma}}
\log\!\frac{1-t}{t}
\right)
+
t\,
\Phi\!\left(
\frac{\sqrt{\Gamma}}{2}
+
\frac{1}{\sqrt{\Gamma}}
\log\!\frac{t}{1-t}
\right),
\]
where $\Phi$ denotes the standard normal distribution function. Independence is approached as $\Gamma\to\infty$, whereas complete dependence corresponds to $\Gamma\to0$. The upper tail dependence coefficient is given by
\begin{equation}
\chi
=
2\left(
1-\Phi\!\left(\frac{\sqrt{\Gamma}}{2}\right)
\right).
\label{eq:chi_husler_reiss}
\end{equation}
Consequently, increasing values of $\Gamma$ lead to weaker extremal dependence.

Another family of extreme-value copulas is the Galambos copula \citep{galambos1975order}. In dimension $p$, its stable tail dependence function is given by
\[
\ell_{\theta}(x_1,\ldots,x_p)
=
\sum_{\varnothing\neq I\subseteq\{1,\ldots,p\}}
(-1)^{|I|+1}
\left(
    \sum_{i\in I}x_i^{-\theta}
\right)^{-1/\theta},
\qquad \theta>0.
\]
Equivalently, the corresponding extreme-value copula is
\[
C_\theta(u_1,\dots,u_p)
=
\exp\!\left\{
-\ell_\theta
\bigl(
-\log u_1,\dots,-\log u_p
\bigr)
\right\}.
\]
The parameter $\theta$ controls the strength of extremal dependence. Independence is recovered when $\theta\to0$, whereas complete dependence is approached as $\theta\to\infty$. In the bivariate case, the stable tail dependence function simplifies to
\[
\ell_\theta(x,y)
=
x+y-
\left(
x^{-\theta}
+
y^{-\theta}
\right)^{-1/\theta},
\]
and the corresponding Pickands dependence function is
\[
A_\theta(t)
=
1-
\left(
t^{-\theta}
+
(1-t)^{-\theta}
\right)^{-1/\theta},
\qquad
t\in[0,1].
\]
The upper tail dependence coefficient admits the closed-form expression
\begin{equation}
\chi
=
2^{-1/\theta}.
\label{eq:chi_galambos}
\end{equation}

Although not belonging to the class of extreme-value copulas, the $t$-copula provides a flexible model for multivariate dependence with non-trivial tail dependence and is widely used in finance and risk management \citep{demarta2005t}. Let $T=(T_1,\dots,T_p)$ follow a multivariate Student-$t$ distribution with correlation matrix $R$ and $\nu>0$ degrees of freedom. The associated $t$-copula is defined by
\[
C_{\nu,R}(u_1,\dots,u_p)
=
t_{\nu,R}
\!\left(
t_\nu^{-1}(u_1),
\dots,
t_\nu^{-1}(u_p)
\right),
\]
where $t_\nu$ denotes the univariate Student-$t$ distribution function with $\nu$ degrees of freedom and $t_{\nu,R}$ the corresponding multivariate distribution function. The dependence structure is governed jointly by the correlation matrix $R$ and the degrees-of-freedom parameter $\nu$. Smaller values of $\nu$ produce heavier tails and stronger extremal dependence. As $\nu\to\infty$, the $t$-copula converges to the Gaussian copula. In the bivariate case, with correlation parameter $\rho$, the upper tail dependence coefficient admits the closed-form expression
\begin{equation}
\chi
=
2\,t_{\nu+1}
\!\left(
-\sqrt{
\frac{(\nu+1)(1-\rho)}
     {1+\rho}
}
\right),
\label{eq:chi_t_copula}
\end{equation}
where $t_{\nu+1}$ denotes the univariate Student-$t$ distribution function with $\nu+1$ degrees of freedom. Consequently, stronger correlation and smaller degrees of freedom lead to stronger extremal dependence.

Unlike the Gumbel--Hougaard, H\"usler--Reiss and Galambos families, the $t$-copula is not itself an extreme-value copula. Nevertheless, it belongs to the max-domain of attraction of the extremal $t$ distribution \citep{opitz2013extremal}. As a result, its asymptotic extremal dependence structure is characterized by an extreme-value copula.

\section{Supplementary material for the method section}
\label{sec:supplementary_method}
\begin{figure}
    \centering
    \includegraphics[scale=.43]{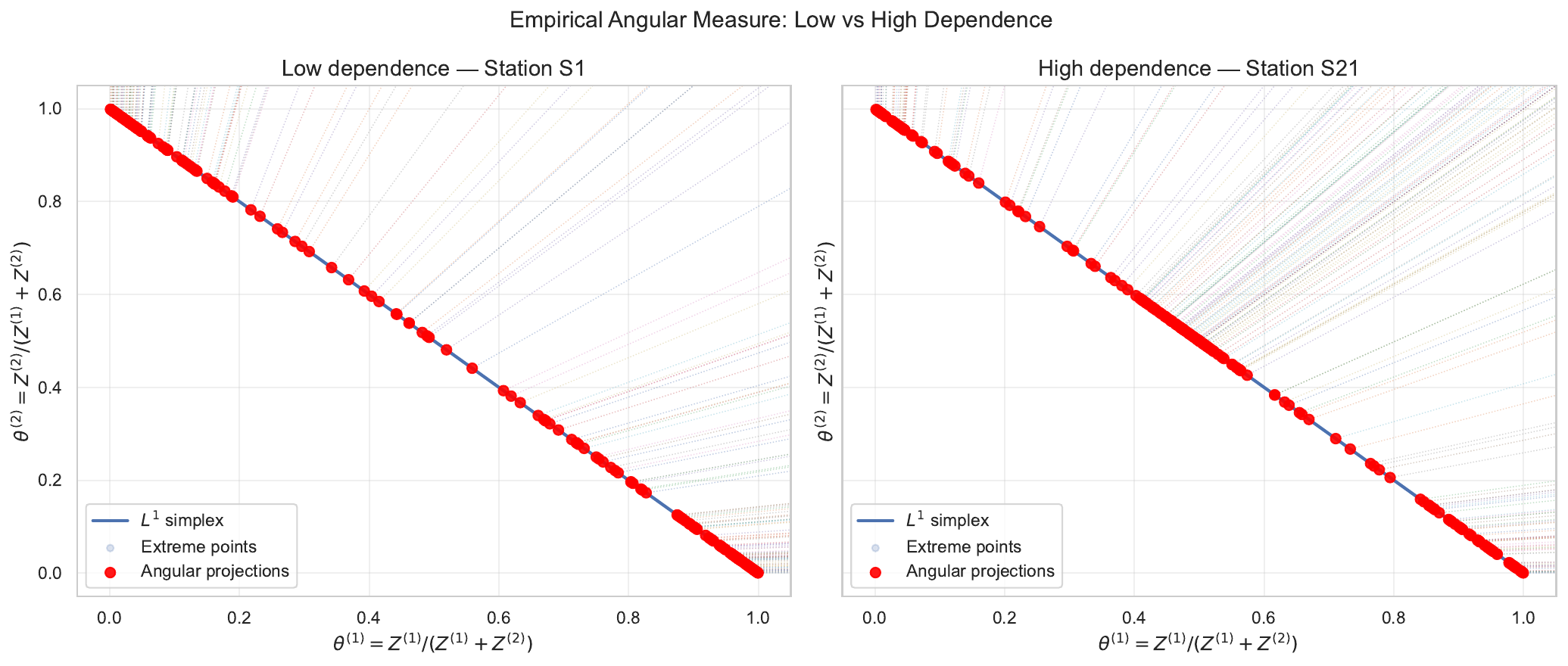}
    \caption{Illustration of angular observations for a bivariate random vector under two extremal dependence regimes.}
    \label{fig:angular_measure_bivariate}
\end{figure}

Figure~\ref{fig:angular_measure_bivariate} illustrate the angular observations for a bivariate random vector on the simplex under two extremal dependence regimes.

\section{Consistency of Wasserstein clustering procedures}
\label{sec:proof_consistency}

This appendix provides the proofs of Lemma~\ref{lem:projected_identifiability} and Theorem~\ref{thm:consistency}.

\begin{proof}[Lemma 1]

If $g(i)=g(l)$, then $G_i=G_l$ by Assumption~\ref{as:groups}. Consequently,
\[
G_i^{(M)}
=
(\Pi_M)_{\#}G_i
=
(\Pi_M)_{\#}G_l
=
G_l^{(M)}
\]
for every $M>0$.
Conversely, suppose that $g(i)\neq g(l)$. By Assumption~\ref{as:groups} and the one-to-one character of the Aitchison-coordinate map,
\[
G_i\neq G_l.
\]
Hence, there exists a Borel set $A\subset\mathbb{R}^{p-1}$ such that
\[
G_i(A)\neq G_l(A).
\]
For $q\geq1$, define
\[
A_q
=
A\cap
\left\{
a\in\mathbb{R}^{p-1}:\|a\|<q
\right\}.
\]
Since $A_q\uparrow A$ as $q\to\infty$, continuity from below gives
\[
G_i(A_q)\longrightarrow G_i(A),
\qquad
G_l(A_q)\longrightarrow G_l(A).
\]
It follows that there exists a finite $q_{il}>0$ such that
\[
G_i(A_{q_{il}})
\neq
G_l(A_{q_{il}}).
\]
For every $M\geq q_{il}$, the set $A_{q_{il}}$ lies strictly inside $\mathcal B_M$. The projection leaves this set unchanged and maps every point outside $\mathcal B_M$ to its boundary. Therefore,
\[
\Pi_M^{-1}(A_{q_{il}})
=
A_{q_{il}},
\]
and hence
\[
G_i^{(M)}(A_{q_{il}})
=
G_i(A_{q_{il}})
\neq
G_l(A_{q_{il}})
=
G_l^{(M)}(A_{q_{il}}).
\]
Thus,
\[
G_i^{(M)}\neq G_l^{(M)}
\]
for every $M\geq q_{il}$.
There are only finitely many between-group pairs. We may therefore choose
\[
M_0
=
\max_{\substack{1\leq i,l\leq n\\g(i)\neq g(l)}}
q_{il}
<
\infty.
\]
For every $M\geq M_0$, we then have
\[
g(i)=g(l)
\quad\Longleftrightarrow\quad
G_i^{(M)}=G_l^{(M)}.
\]
If $K\geq2$, the set of between-group pairs is finite and nonempty. Moreover, the projected measures are supported on $\mathcal B_M$ and therefore possess finite first moments. Since $W_1$ is a metric on this space,
\[
W_1\left(G_i^{(M)},G_l^{(M)}\right)>0
\]
for every between-group pair. Consequently,
\[
\Delta_M
=
\min_{g(i)\neq g(l)}
W_1\left(G_i^{(M)},G_l^{(M)}\right)
>
0.
\]
\end{proof}

\begin{proof}[ Theorem~\ref{thm:consistency}]
If $K=1$, both clustering procedures return the unique partition containing all random vectors, and the result is immediate. We therefore assume throughout the proof that $K\geq2$ and fix $M\geq M_0$.

\medskip
\noindent
\textit{Step 1: Wasserstein consistency of the projected empirical measures.}

Under Assumptions~\ref{as:limit} and~\ref{as:threshold}, together with the sampling conditions stated in the main text, the empirical angular measures satisfy
\[
\widehat H_{i,T}
\rightsquigarrow
H_i
\]
in probability for every $i\in\{1,\dots,n\}$; see the proof of Proposition~3.3 of \citet{janssen2020kmeans}.
The composition
\[
\Pi_M\circ\mathcal A:
\Delta_{p-1}^{\circ}
\longrightarrow
\mathcal B_M
\]
is continuous. The Continuous Mapping Theorem therefore gives
\[
\begin{split}
\widehat G_{i,T}^{(M)}
&=
(\Pi_M)_{\#}\widehat G_{i,T}\\
&=
(\Pi_M\circ\mathcal A)_{\#}\widehat H_{i,T}
\rightsquigarrow
(\Pi_M\circ\mathcal A)_{\#}H_i\\
&=
(\Pi_M)_{\#}G_i
=
G_i^{(M)}
\end{split}
\]
in probability. To verify convergence of the first moments, define
\[
f_M(a)
=
\min\{\|a\|,M\},
\qquad
a\in\mathbb{R}^{p-1}.
\]
The function $f_M$ is bounded and continuous. Since both projected measures are supported on $\mathcal B_M$,
\[
\begin{split}
\int_{\mathbb{R}^{p-1}}
\|a\|
\,\mathrm{d}\widehat G_{i,T}^{(M)}(a)
&=
\int_{\mathbb{R}^{p-1}}
f_M(a)
\,\mathrm{d}\widehat G_{i,T}^{(M)}(a)\\
&\longrightarrow
\int_{\mathbb{R}^{p-1}}
f_M(a)
\,\mathrm{d}G_i^{(M)}(a)\\
&=
\int_{\mathbb{R}^{p-1}}
\|a\|
\,\mathrm{d}G_i^{(M)}(a)
\end{split}
\]
in probability. Thus, weak convergence and convergence of the first moments hold. By Theorem~6.9 of \citet{villani2009optimal},
\[
W_1\left(
\widehat G_{i,T}^{(M)},
G_i^{(M)}
\right)
\longrightarrow
0
\]
in probability for every $i$.

\medskip
\noindent
\textit{Step 2: Uniform consistency of the empirical dissimilarity matrix.}

Define the population projected dissimilarities by
\[
d_{il}^{W,M}
=
W_1\left(
G_i^{(M)},
G_l^{(M)}
\right).
\]
For every $i,l\in\{1,\dots,n\}$, the triangle inequality gives
\[
|
\widehat d_{il,T}^{W,M}
-
d_{il}^{W,M}
|
\leq
W_1\left(
\widehat G_{i,T}^{(M)},
G_i^{(M)}
\right)
+
W_1\left(
\widehat G_{l,T}^{(M)},
G_l^{(M)}
\right).
\]
The two terms on the right-hand side converge to zero in probability. Since $n$ is fixed, it follows that
\[
\varepsilon_T
:=
\max_{1\leq i,l\leq n}
|
\widehat d_{il,T}^{W,M}
-
d_{il}^{W,M}
|
\longrightarrow
0
\]
in probability. By Lemma~\ref{lem:projected_identifiability},
\[
d_{il}^{W,M}=0
\quad\text{if }g(i)=g(l),
\]
whereas
\[
d_{il}^{W,M}\geq\Delta_M
\quad\text{if }g(i)\neq g(l).
\]
Define
\[
A_T
=
\left\{
\varepsilon_T
<
\frac{\Delta_M}{3n}
\right\}.
\]
The preceding uniform convergence implies that
\[
\operatorname{pr}(A_T)\longrightarrow 1.
\]
On $A_T$, every within-group empirical dissimilarity satisfies
\[
\widehat d_{il,T}^{W,M}
\leq
\varepsilon_T
<
\frac{\Delta_M}{3n},
\qquad
g(i)=g(l),
\]
whereas every between-group empirical dissimilarity satisfies
\[
\widehat d_{il,T}^{W,M}
\geq
\Delta_M-\varepsilon_T,
\qquad
g(i)\neq g(l).
\]
Since $\varepsilon_T<\Delta_M/2$ on $A_T$, every within-group empirical dissimilarity is strictly smaller than every between-group empirical dissimilarity.

\medskip
\noindent
\textit{Step 3: Consistency of hierarchical clustering.}

Initially, every cluster is a singleton and is therefore contained in one true latent group. Suppose that, at some stage of the hierarchical procedure, every cluster formed so far remains contained in a single true group.

Let $C$ and $C'$ be two current clusters. If they are contained in the same true group, all pairwise dissimilarities between their elements are bounded above by $\varepsilon_T$. Their single-, complete-, and average-linkage dissimilarities are therefore also bounded above by $\varepsilon_T$.

If $C$ and $C'$ are contained in different true groups, all pairwise dissimilarities between their elements are bounded below by $\Delta_M-\varepsilon_T$. Their single-, complete-, and average-linkage dissimilarities are therefore also bounded below by $\Delta_M-\varepsilon_T$.

As long as more than $K$ clusters remain, at least one true group is split across two or more current clusters. Hence, at least one within-group merge is available. On $A_T$,
\[
\varepsilon_T
<
\Delta_M-\varepsilon_T,
\]
so every available within-group merge has a smaller linkage dissimilarity than every between-group merge. The next merge must therefore occur within a true group.

By induction, no between-group merge occurs before every true group has been merged into a single cluster. Consequently, when the dendrogram is cut at exactly $K$ clusters, the resulting partition is $\mathcal P$.

\medskip
\noindent
\textit{Step 4: Consistency of the global $K$-medoids solution.}

Write the true partition as
\[
\mathcal P
=
\left\{
\mathcal P_1,\dots,\mathcal P_K
\right\},
\qquad
\mathcal P_k
=
\left\{
i\in\{1,\dots,n\}:g(i)=k
\right\}.
\]
For any set $\mathcal S\subseteq\{1,\dots,n\}$ containing exactly $K$ medoid indices, define the population and empirical $K$-medoids objectives by
\[
\Phi^{(M)}(\mathcal S)
=
\sum_{i=1}^n
\min_{s\in\mathcal S}
d_{is}^{W,M}
\]
and
\[
\widehat\Phi_T^{(M)}(\mathcal S)
=
\sum_{i=1}^n
\min_{s\in\mathcal S}
\widehat d_{is,T}^{W,M}.
\]

Suppose first that $\mathcal S$ contains exactly one index from each true group. Every $i\in\{1,\dots,n\}$ then has a medoid $s\in\mathcal S$ belonging to the same true group. Consequently,
\[
\Phi^{(M)}(\mathcal S)=0.
\]

Conversely, suppose that $\mathcal S$ does not contain an index from every true group. Since $\mathcal S$ contains $K$ indices and there are $K$ true groups, at least one true group, say $\mathcal P_k$, is not represented. For every $i\in\mathcal P_k$ and every $s\in\mathcal S$,
\[
d_{is}^{W,M}\geq\Delta_M.
\]
Since $\mathcal P_k$ is nonempty, this implies
\[
\Phi^{(M)}(\mathcal S)\geq\Delta_M.
\]
Thus, the population objective is zero precisely for those medoid sets containing one representative from each true group. Although the medoid set need not be unique, every population minimizer represents all $K$ groups. For every admissible medoid set $\mathcal S$,
\[
\begin{split}
|
\widehat\Phi_T^{(M)}(\mathcal S)
-
\Phi^{(M)}(\mathcal S)
|
&\leq
\sum_{i=1}^n
|
\min_{s\in\mathcal S}
\widehat d_{is,T}^{W,M}
-
\min_{s\in\mathcal S}
d_{is}^{W,M}
|\\
&\leq
n\varepsilon_T.
\end{split}
\]
On $A_T$, we have $n\varepsilon_T<\Delta_M/3$. Hence, if $\mathcal S$ contains one medoid from each true group,
\[
\widehat\Phi_T^{(M)}(\mathcal S)
<
\frac{\Delta_M}{3}.
\]
If $\mathcal S$ omits at least one true group, then
\[
\widehat\Phi_T^{(M)}(\mathcal S)
>
\Delta_M-\frac{\Delta_M}{3}
=
\frac{2\Delta_M}{3}.
\]
Therefore, on $A_T$, every global minimizer $\widehat{\mathcal S}_T^{(M)}$ of the empirical $K$-medoids objective contains exactly one medoid from each true group.

It remains to consider the assignment to the selected medoids. Let $i\in\mathcal P_k$, and let $s_k$ denote the selected medoid belonging to $\mathcal P_k$. Then
\[
\widehat d_{is_k,T}^{W,M}
\leq
\varepsilon_T.
\]
For every selected medoid $s\notin\mathcal P_k$,
\[
\widehat d_{is,T}^{W,M}
\geq
\Delta_M-\varepsilon_T.
\]
On $A_T$, $\varepsilon_T<\Delta_M-\varepsilon_T$. Thus, every random vector is strictly closer to the selected medoid from its own true group than to any medoid from another group. The nearest-medoid assignment therefore recovers $\mathcal P$ exactly.

We have shown that, on $A_T$, both clustering procedures appearing in Theorem~\ref{thm:consistency} recover the true partition:
\[
\widehat{\mathcal P}_T^{(M)}
=
\mathcal P.
\]
Consequently,
\[
\operatorname{pr}\left(
\widehat{\mathcal P}_T^{(M)}
=
\mathcal P
\right)
\geq
\operatorname{pr}(A_T).
\]
Since $\operatorname{pr}(A_T)\to 1$, it follows that
\[
\operatorname{pr}\left(
\widehat{\mathcal P}_T^{(M)}
=
\mathcal P
\right)
\longrightarrow
1,
\qquad
T\to\infty.
\]
\end{proof}

\section{Supplementary for the simulation studies}\label{sec:supp_simulation}

\subsection{Competing methods}
\label{sec:Competitors}
\citet{vignotto2021clustering} proposed a clustering methodology for compound extremes based on a dissimilarity measure derived from the Kullback--Leibler divergence. The methodology first transforms each marginal distribution to the standard Pareto scale as in Equation~\eqref{eq:pareto_transformation}. To identify extreme regions, they introduce a risk function $r:\Rset^2\to\Rset$, applied to the Pareto-transformed observations. Typical choices include
\[
r(z)=\max\left(z^{(1)},z^{(2)}\right),
\qquad
r(z)=z^{(1)}+z^{(2)}.
\]
For each random vector $i$, they define $Z_{i,t}=\left(Z_{i,t}^{(1)},Z_{i,t}^{(2)}\right)$ and the associated univariate random variables $R_{i,t}=r(Z_{i,t})$. A point is considered extreme whenever $R_{i,t}>q_{u,i}$, where $q_{u,i}$ denotes a high quantile associated with an exceedance probability $u\in(0,1)$. The corresponding tail region is therefore $\left\{R_{i,t}>q_{u,i}\right\}$. They proposed to partition the extreme region into $W$ disjoint subsets $A_1,\ldots,A_W$. The empirical proportion of extreme observations falling into region $A_w$ is estimated by
\[
\widehat p_{i,w}
=
\frac{
\sum_{t=1}^T
\ind\left\{
Z_{i,t}\in A_w
\right\}
}{
\sum_{t=1}^T
\ind\left\{
R_{i,t}>q_{u,i}
\right\}
},
\qquad
w=1,\ldots,W.
\]
The vector $(\widehat p_{i,1},\ldots,\widehat p_{i,W})$ can therefore be interpreted as an empirical multinomial distribution describing how extremes are distributed across the tail regions. The dissimilarity between the extremal dependence structures of two random vectors $i$ and $l$ is then quantified using the symmetrized Kullback--Leibler divergence \citep{van2014renyi},
\[
\widehat d_{il}^{\mathrm{KL}}
=
\frac{1}{2}
\sum_{w=1}^W
\left(
\widehat p_{i,w}-\widehat p_{l,w}
\right)
\log
\left(
\frac{\widehat p_{i,w}}{\widehat p_{l,w}}
\right).
\]
A small value of $\widehat d_{il}^{\mathrm{KL}}$ indicates that both random vectors exhibit similar extremal dependence structures, whereas large values indicate different tail behaviours. Finally, the pairwise divergences are assembled into a dissimilarity matrix
\begin{equation}
\widehat\bD^{\mathrm{KL}}_T
=
\left(
\widehat d_{il}^{\mathrm{KL}}
\right)_{i,l=1,\ldots,n}
\in\Rset^{n\times n},
\label{eq:Vignotto_D_mat}
\end{equation}
which is then used as input for a clustering algorithm. The parameter $W$ controls the resolution of the tail partition. Large values of $W$ allow finer differences in the extremal dependence structure to be detected, but may lead to unstable probability estimates when too few observations fall into each subset. The choice of $W$ therefore induces a classical bias--variance trade-off.

We consider another competing approach, conditional extremes clustering, which was introduced more recently by \citet{o2025clustering}. The method is based on the conditional extremes framework \citep{heffernan2004conditional}, which models the distribution of a random vector conditional on one of its components being extreme. Separate conditional extremes models are fitted for each random vector, yielding parameters that characterize its extremal dependence structure. To compare extremal dependence across random vectors, \citet{o2025clustering} introduce a dissimilarity measure based on the expected skew-geometric Jensen--Shannon divergence \citep{menendez1997jensen} between the fitted conditional distributions. Under a Gaussian working assumption for the conditional extremes residuals, this divergence admits a closed-form expression, allowing efficient computation of pairwise dissimilarities. For multivariate observations, the procedure is repeated for each conditioning component $j\in\{1,\ldots,p\}$. Let $\widehat{\bD}^{(j),\mathrm{CE}}_T$ denote the dissimilarity matrix obtained by fitting the conditional extremes model with component $j$ as the conditioning variable. The final dissimilarity matrix is then constructed by averaging over all conditioning components,
\begin{equation}
\widehat{\bD}^{\mathrm{CE}}_T
=
\frac{1}{p}
\sum_{j=1}^{p}
\widehat{\bD}^{(j),\mathrm{CE}}_T.
\label{eq:CeCl_D_mat}
\end{equation}
This aggregation yields a single dissimilarity matrix that incorporates information from all possible conditioning variables.

\subsection{Evaluation metric}
\label{sec:supp_simulation_ARI}
To evaluate the performance of the clustering in the simulation study, we use the Adjusted Rand Index (ARI), which measures the agreement between the true partition and the partition estimated by the clustering algorithm \citep{hubert1985comparing}. The ARI is a corrected version of the Rand Index that accounts for agreement occurring by chance.
Let $\mathcal P=\left\{P_1,\dots,P_K\right\}$ denote the true partition of the random vectors, and let $\widehat{\mathcal P}=\left\{\widehat{P}_1,\dots,\widehat{P}_{\widehat{K}}\right\}$ be the partition estimated by the clustering procedure. The ARI is based on all pairwise comparisons between random vectors. More precisely, it compares whether two random vectors belong to the same cluster or to different clusters under both partitions. Let $n_{k\ell}=|P_k\cap\widehat{P}_\ell|$ denote the contingency table associated with the two partitions, and define
\[
a_k
=
\sum_{\ell=1}^{\widehat{K}}
n_{k\ell},
\qquad
b_\ell
=
\sum_{k=1}^{K}
n_{k\ell}.
\]
The Adjusted Rand Index is then defined by \citep{hubert1985comparing,steinley2004properties}
\begin{equation}
\mathrm{ARI}
=
\frac{
\displaystyle
\sum_{k,\ell}
\binom{n_{k\ell}}{2}
-
\frac{
\displaystyle
\sum_k \binom{a_k}{2}
\sum_\ell \binom{b_\ell}{2}
}{
\binom{n}{2}
}
}{
\displaystyle
\frac12
\left[
\sum_k \binom{a_k}{2}
+
\sum_\ell \binom{b_\ell}{2}
\right]
-
\frac{
\displaystyle
\sum_k \binom{a_k}{2}
\sum_\ell \binom{b_\ell}{2}
}{
\binom{n}{2}
}
}.
\label{eq:ARI}
\end{equation}
The ARI takes values in $[-1,1]$. A value close to $1$ indicates perfect agreement between the true and estimated partitions, whereas values close to $0$ correspond to a clustering performance similar to random assignment. Negative values indicate agreement worse than random clustering.

\subsection{Detailed Results for Simulation Study 1}
\label{sec:supp_simulation_tables}
% \label{sec:supp_simulation_results}
\begin{figure}
    \centering
    \includegraphics[
        width=\textwidth,
        height=0.9\textheight,
        keepaspectratio
    ]{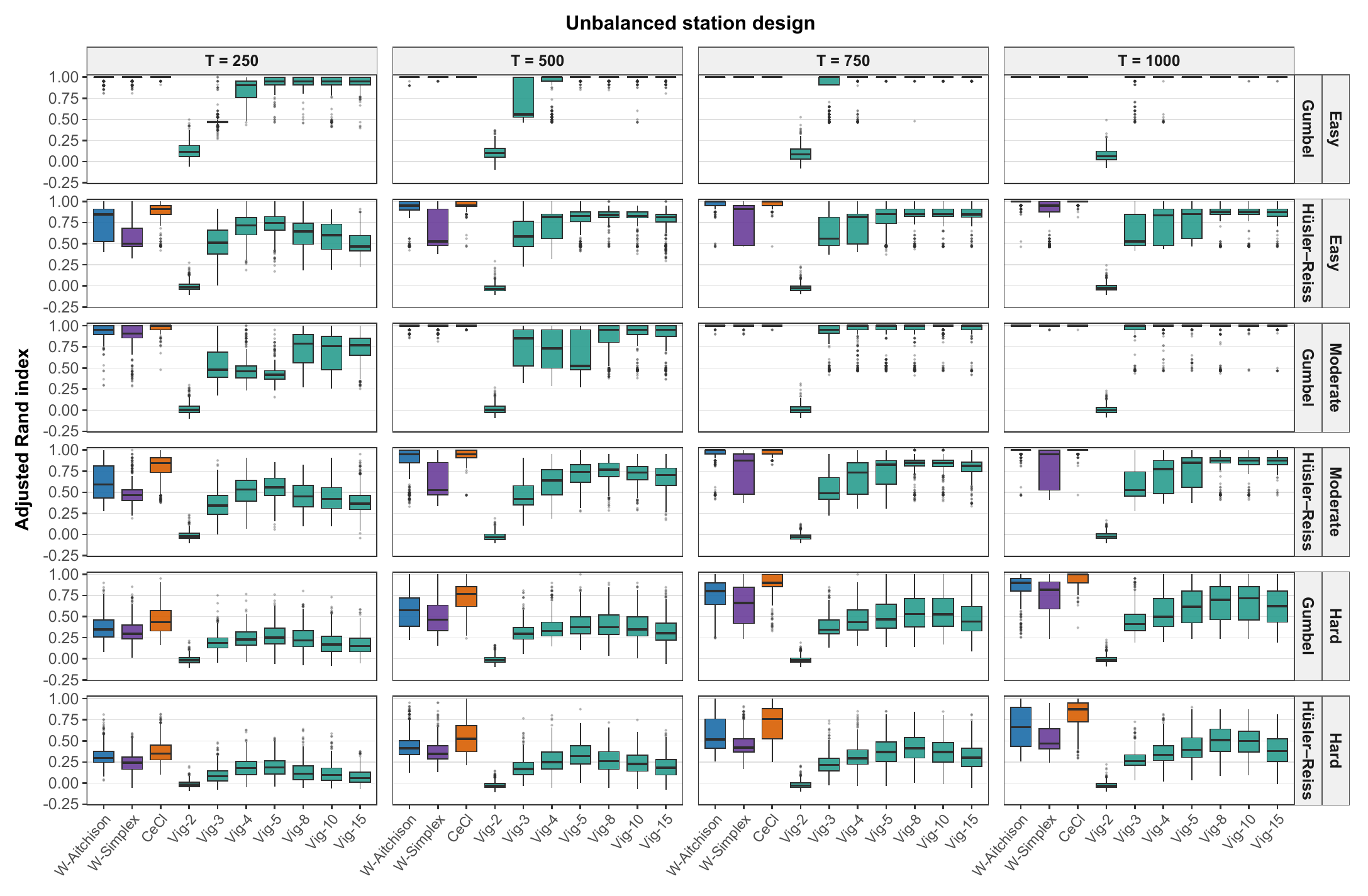}
    \caption{Adjusted Rand index for the unbalanced design $N=(5,10,20)$, with $K=3$ latent groups, for sample sizes $T\in\{250,500,750,\num{1000}\}$ and the three separation levels easy, moderate and hard, corresponding to $\chi=(0.1,0.5,0.9)$, $(0.2,0.5,0.8)$ and $(0.3,0.45,0.6)$, under the Gumbel--Hougaard and H\"usler--Reiss copulas. The Wasserstein--Aitchison, Wasserstein--simplex and conditional extremes dissimilarities are compared with the partition-based dissimilarity of \citet{vignotto2021clustering}, the latter for $W\in\{2,3,4,5,8,10,15\}$ partitions. Results are based on 500 Monte Carlo replications.}
    \label{fig:ari_unbalanced}
\end{figure}

\begin{figure}%[p]
    \centering
    \includegraphics[
        width=\textwidth,
        height=0.9\textheight,
        keepaspectratio]{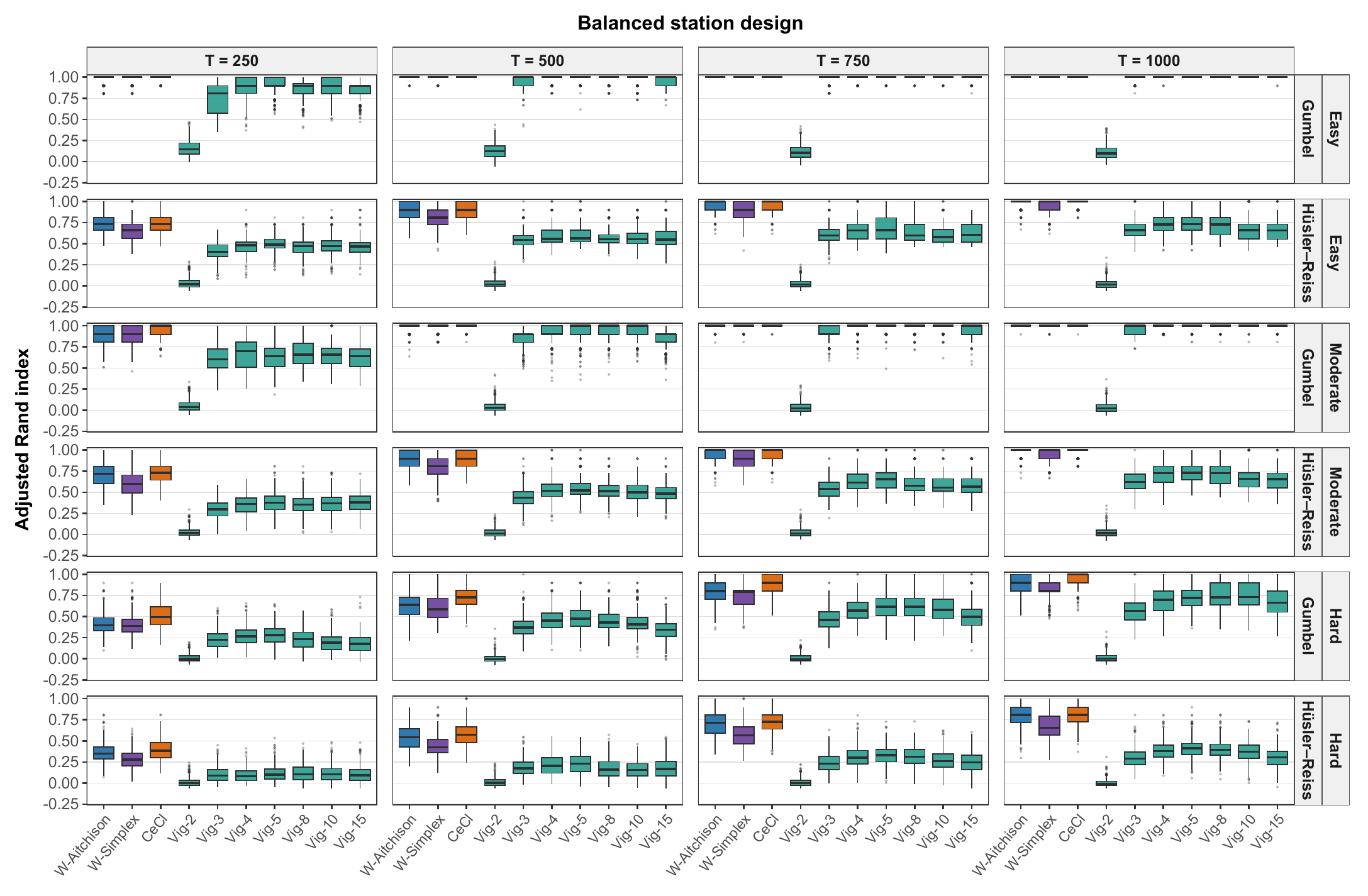}
    \caption{ Adjusted Rand Index (ARI) for the balanced design $N=(10,\,10,\,10)$, with $K=3$ latent groups, for sample sizes $T\in\{250,\,500,\,750,\,\num{1000}\}$ and the three separation levels easy, moderate and hard ($\chi=(0.1,\,0.5,\,0.9)$, $(0.2,\,0.5,\,0.8)$ and $(0.3,\,0.45,\,0.6)$), under the Gumbel--Hougaard and H\"usler--Reiss copulas. The Wasserstein--Aitchison, Wasserstein--simplex and CeCl dissimilarities are compared with the partition-based dissimilarity of \citet{vignotto2021clustering}, the latter for $W\in\{2,\,3,\,4,\,5,\,8,\,10,\,15\}$ partitions. Results are based on 500 Monte Carlo replications.}
    \label{fig:ari_balanced}
\end{figure}

Table~\ref{tab:simulation1_mean_ari} reports the corresponding mean values
and Monte Carlo standard errors. Table~
\ref{tab:simulation1_exact_recovery} reports the percentage of replications
in which the latent partition is recovered exactly.

\begingroup\fontsize{7}{9}\selectfont

\begin{longtable}[t]{lrrrrrrr}
\caption{\label{tab:simulation1_mean_ari}Mean adjusted Rand index, with Monte Carlo standard errors in parentheses, based on 500 replications. WA and WS denote Wasserstein--Aitchison and Wasserstein--simplex. The Vignotto column is an oracle comparison obtained by selecting the value of $W$ with the largest mean adjusted Rand index separately within each simulation setting.}\\
\toprule
Design & Copula & Scenario & $T$ & WA & WS & CeCl & Vignotto\\
\midrule
\endfirsthead
\caption[]{Mean adjusted Rand index, with Monte Carlo standard errors in parentheses, based on 500 replications. WA and WS denote Wasserstein--Aitchison and Wasserstein--simplex. The Vignotto column is an oracle comparison obtained by selecting the value of $W$ with the largest mean adjusted Rand index separately within each simulation setting. \textit{(continued)}}\\
\toprule
Design & Copula & Scenario & $T$ & WA & WS & CeCl & Vignotto\\
\midrule
\endhead

\endfoot
\bottomrule
\endlastfoot
Balanced & Gumbel & Easy & \num{250} & 0.978 (0.002) & 0.978 (0.002) & 0.996 (0.001) & 0.906 (0.005) [$W=5$]\\
Balanced & Gumbel & Easy & \num{500} & 0.999 (0.000) & 0.998 (0.001) & 1.000 (0.000) & 0.991 (0.001) [$W=5$]\\
Balanced & Gumbel & Easy & \num{750} & 1.000 (0.000) & 1.000 (0.000) & 1.000 (0.000) & 0.999 (0.001) [$W=8$]\\
Balanced & Gumbel & Easy & \num{1000} & 1.000 (0.000) & 1.000 (0.000) & 1.000 (0.000) & 1.000 (0.000) [$W=5$]\\
Balanced & Gumbel & Moderate & \num{250} & 0.880 (0.005) & 0.880 (0.005) & 0.950 (0.003) & 0.667 (0.006) [$W=8$]\\
\addlinespace
Balanced & Gumbel & Moderate & \num{500} & 0.983 (0.002) & 0.980 (0.002) & 0.996 (0.001) & 0.941 (0.004) [$W=8$]\\
Balanced & Gumbel & Moderate & \num{750} & 0.997 (0.001) & 0.997 (0.001) & 0.999 (0.000) & 0.985 (0.002) [$W=10$]\\
Balanced & Gumbel & Moderate & \num{1000} & 1.000 (0.000) & 0.999 (0.000) & 1.000 (0.000) & 0.999 (0.001) [$W=10$]\\
Balanced & Gumbel & Hard & \num{250} & 0.416 (0.006) & 0.396 (0.006) & 0.510 (0.006) & 0.278 (0.005) [$W=5$]\\
Balanced & Gumbel & Hard & \num{500} & 0.635 (0.006) & 0.597 (0.006) & 0.744 (0.006) & 0.490 (0.006) [$W=5$]\\
\addlinespace
Balanced & Gumbel & Hard & \num{750} & 0.774 (0.006) & 0.753 (0.006) & 0.873 (0.005) & 0.618 (0.007) [$W=8$]\\
Balanced & Gumbel & Hard & \num{1000} & 0.862 (0.005) & 0.835 (0.005) & 0.934 (0.004) & 0.743 (0.007) [$W=8$]\\
Balanced & H{\"u}sler--Reiss & Easy & \num{250} & 0.751 (0.005) & 0.660 (0.005) & 0.755 (0.005) & 0.494 (0.004) [$W=5$]\\
Balanced & H{\"u}sler--Reiss & Easy & \num{500} & 0.896 (0.004) & 0.809 (0.005) & 0.898 (0.004) & 0.604 (0.005) [$W=5$]\\
Balanced & H{\"u}sler--Reiss & Easy & \num{750} & 0.954 (0.003) & 0.884 (0.005) & 0.957 (0.003) & 0.678 (0.006) [$W=5$]\\
\addlinespace
Balanced & H{\"u}sler--Reiss & Easy & \num{1000} & 0.978 (0.002) & 0.938 (0.004) & 0.982 (0.002) & 0.748 (0.006) [$W=5$]\\
Balanced & H{\"u}sler--Reiss & Moderate & \num{250} & 0.705 (0.006) & 0.601 (0.006) & 0.734 (0.006) & 0.379 (0.005) [$W=5$]\\
Balanced & H{\"u}sler--Reiss & Moderate & \num{500} & 0.890 (0.004) & 0.786 (0.006) & 0.896 (0.004) & 0.544 (0.005) [$W=5$]\\
Balanced & H{\"u}sler--Reiss & Moderate & \num{750} & 0.944 (0.003) & 0.888 (0.005) & 0.954 (0.003) & 0.658 (0.006) [$W=5$]\\
Balanced & H{\"u}sler--Reiss & Moderate & \num{1000} & 0.983 (0.002) & 0.932 (0.004) & 0.982 (0.002) & 0.743 (0.006) [$W=5$]\\
\addlinespace
Balanced & H{\"u}sler--Reiss & Hard & \num{250} & 0.365 (0.005) & 0.282 (0.005) & 0.395 (0.005) & 0.117 (0.005) [$W=8$]\\
Balanced & H{\"u}sler--Reiss & Hard & \num{500} & 0.551 (0.006) & 0.442 (0.006) & 0.583 (0.006) & 0.226 (0.005) [$W=5$]\\
Balanced & H{\"u}sler--Reiss & Hard & \num{750} & 0.696 (0.006) & 0.574 (0.006) & 0.728 (0.006) & 0.326 (0.005) [$W=5$]\\
Balanced & H{\"u}sler--Reiss & Hard & \num{1000} & 0.785 (0.006) & 0.663 (0.007) & 0.816 (0.006) & 0.414 (0.005) [$W=5$]\\
Unbalanced & Gumbel & Easy & \num{250} & 0.986 (0.002) & 0.985 (0.001) & 0.999 (0.000) & 0.940 (0.003) [$W=8$]\\
\addlinespace
Unbalanced & Gumbel & Easy & \num{500} & 0.999 (0.000) & 0.999 (0.000) & 1.000 (0.000) & 0.996 (0.001) [$W=5$]\\
Unbalanced & Gumbel & Easy & \num{750} & 1.000 (0.000) & 1.000 (0.000) & 1.000 (0.000) & 0.999 (0.000) [$W=5$]\\
Unbalanced & Gumbel & Easy & \num{1000} & 1.000 (0.000) & 1.000 (0.000) & 1.000 (0.000) & 1.000 (0.000) [$W=5$]\\
Unbalanced & Gumbel & Moderate & \num{250} & 0.924 (0.004) & 0.881 (0.007) & 0.974 (0.002) & 0.734 (0.008) [$W=8$]\\
Unbalanced & Gumbel & Moderate & \num{500} & 0.991 (0.001) & 0.990 (0.001) & 0.999 (0.000) & 0.896 (0.007) [$W=10$]\\
\addlinespace
Unbalanced & Gumbel & Moderate & \num{750} & 0.999 (0.000) & 0.999 (0.000) & 1.000 (0.000) & 0.974 (0.004) [$W=10$]\\
Unbalanced & Gumbel & Moderate & \num{1000} & 1.000 (0.000) & 1.000 (0.000) & 1.000 (0.000) & 0.995 (0.002) [$W=10$]\\
Unbalanced & Gumbel & Hard & \num{250} & 0.376 (0.007) & 0.327 (0.006) & 0.464 (0.007) & 0.271 (0.006) [$W=5$]\\
Unbalanced & Gumbel & Hard & \num{500} & 0.566 (0.009) & 0.493 (0.009) & 0.730 (0.008) & 0.411 (0.007) [$W=8$]\\
Unbalanced & Gumbel & Hard & \num{750} & 0.739 (0.009) & 0.634 (0.010) & 0.887 (0.006) & 0.550 (0.009) [$W=10$]\\
\addlinespace
Unbalanced & Gumbel & Hard & \num{1000} & 0.848 (0.007) & 0.752 (0.010) & 0.946 (0.003) & 0.666 (0.010) [$W=8$]\\
Unbalanced & H{\"u}sler--Reiss & Easy & \num{250} & 0.770 (0.009) & 0.581 (0.007) & 0.875 (0.005) & 0.725 (0.006) [$W=5$]\\
Unbalanced & H{\"u}sler--Reiss & Easy & \num{500} & 0.899 (0.007) & 0.659 (0.009) & 0.954 (0.003) & 0.834 (0.004) [$W=8$]\\
Unbalanced & H{\"u}sler--Reiss & Easy & \num{750} & 0.964 (0.003) & 0.759 (0.010) & 0.979 (0.002) & 0.850 (0.004) [$W=10$]\\
Unbalanced & H{\"u}sler--Reiss & Easy & \num{1000} & 0.986 (0.002) & 0.860 (0.009) & 0.992 (0.001) & 0.866 (0.004) [$W=15$]\\
\addlinespace
Unbalanced & H{\"u}sler--Reiss & Moderate & \num{250} & 0.627 (0.009) & 0.495 (0.006) & 0.798 (0.008) & 0.555 (0.007) [$W=5$]\\
Unbalanced & H{\"u}sler--Reiss & Moderate & \num{500} & 0.858 (0.009) & 0.633 (0.009) & 0.945 (0.003) & 0.749 (0.005) [$W=8$]\\
Unbalanced & H{\"u}sler--Reiss & Moderate & \num{750} & 0.933 (0.006) & 0.747 (0.010) & 0.980 (0.002) & 0.831 (0.005) [$W=8$]\\
Unbalanced & H{\"u}sler--Reiss & Moderate & \num{1000} & 0.981 (0.003) & 0.831 (0.010) & 0.991 (0.001) & 0.865 (0.003) [$W=15$]\\
Unbalanced & H{\"u}sler--Reiss & Hard & \num{250} & 0.321 (0.005) & 0.240 (0.005) & 0.377 (0.006) & 0.193 (0.005) [$W=5$]\\
\addlinespace
Unbalanced & H{\"u}sler--Reiss & Hard & \num{500} & 0.447 (0.007) & 0.377 (0.006) & 0.542 (0.008) & 0.335 (0.007) [$W=5$]\\
Unbalanced & H{\"u}sler--Reiss & Hard & \num{750} & 0.574 (0.009) & 0.456 (0.006) & 0.706 (0.009) & 0.424 (0.007) [$W=8$]\\
Unbalanced & H{\"u}sler--Reiss & Hard & \num{1000} & 0.662 (0.010) & 0.524 (0.008) & 0.802 (0.009) & 0.507 (0.008) [$W=8$]\\*
\end{longtable}
\endgroup{}

\begingroup\fontsize{7}{9}\selectfont

\begin{longtable}[t]{lrrrrrrr}
\caption{\label{tab:simulation1_exact_recovery}Percentage of the 500 replications achieving exact recovery, $\operatorname{ARI}=1$. WA and WS denote Wasserstein--Aitchison and Wasserstein--simplex. For Vignotto, $W$ is selected using the mean adjusted Rand index reported in Table~\ref{tab:simulation1_mean_ari}.}\\
\toprule
Design & Copula & Scenario & $T$ & WA & WS & CeCl & Vignotto\\
\midrule
\endfirsthead
\caption[]{Percentage of the 500 replications achieving exact recovery, $\operatorname{ARI}=1$. WA and WS denote Wasserstein--Aitchison and Wasserstein--simplex. For Vignotto, $W$ is selected using the mean adjusted Rand index reported in Table~\ref{tab:simulation1_mean_ari}. \textit{(continued)}}\\
\toprule
Design & Copula & Scenario & $T$ & WA & WS & CeCl & Vignotto\\
\midrule
\endhead

\endfoot
\bottomrule
\endlastfoot
Balanced & Gumbel & Easy & \num{250} & 80.0\% & 79.6\% & 96.0\% & 41.8\% [$W=5$]\\
Balanced & Gumbel & Easy & \num{500} & 98.8\% & 97.8\% & 100.0\% & 92.2\% [$W=5$]\\
Balanced & Gumbel & Easy & \num{750} & 100.0\% & 100.0\% & 100.0\% & 98.6\% [$W=8$]\\
Balanced & Gumbel & Easy & \num{1000} & 100.0\% & 100.0\% & 100.0\% & 100.0\% [$W=5$]\\
Balanced & Gumbel & Moderate & \num{250} & 32.2\% & 30.2\% & 61.8\% & 3.4\% [$W=8$]\\
\addlinespace
Balanced & Gumbel & Moderate & \num{500} & 85.2\% & 83.0\% & 96.0\% & 58.4\% [$W=8$]\\
Balanced & Gumbel & Moderate & \num{750} & 97.2\% & 96.8\% & 99.4\% & 87.2\% [$W=10$]\\
Balanced & Gumbel & Moderate & \num{1000} & 99.8\% & 99.0\% & 99.8\% & 99.0\% [$W=10$]\\
Balanced & Gumbel & Hard & \num{250} & 0.0\% & 0.0\% & 0.0\% & 0.0\% [$W=5$]\\
Balanced & Gumbel & Hard & \num{500} & 0.6\% & 1.0\% & 4.4\% & 0.0\% [$W=5$]\\
\addlinespace
Balanced & Gumbel & Hard & \num{750} & 7.8\% & 6.2\% & 28.4\% & 1.4\% [$W=8$]\\
Balanced & Gumbel & Hard & \num{1000} & 26.2\% & 18.0\% & 52.6\% & 7.8\% [$W=8$]\\
Balanced & H{\"u}sler--Reiss & Easy & \num{250} & 3.6\% & 0.8\% & 4.4\% & 0.0\% [$W=5$]\\
Balanced & H{\"u}sler--Reiss & Easy & \num{500} & 34.6\% & 12.2\% & 36.8\% & 1.0\% [$W=5$]\\
Balanced & H{\"u}sler--Reiss & Easy & \num{750} & 64.4\% & 31.4\% & 67.2\% & 2.0\% [$W=5$]\\
\addlinespace
Balanced & H{\"u}sler--Reiss & Easy & \num{1000} & 81.4\% & 54.6\% & 82.8\% & 6.8\% [$W=5$]\\
Balanced & H{\"u}sler--Reiss & Moderate & \num{250} & 2.6\% & 0.2\% & 4.0\% & 0.0\% [$W=5$]\\
Balanced & H{\"u}sler--Reiss & Moderate & \num{500} & 31.4\% & 9.4\% & 36.2\% & 0.4\% [$W=5$]\\
Balanced & H{\"u}sler--Reiss & Moderate & \num{750} & 57.6\% & 34.4\% & 64.8\% & 1.2\% [$W=5$]\\
Balanced & H{\"u}sler--Reiss & Moderate & \num{1000} & 85.6\% & 52.2\% & 83.8\% & 6.6\% [$W=5$]\\
\addlinespace
Balanced & H{\"u}sler--Reiss & Hard & \num{250} & 0.0\% & 0.0\% & 0.0\% & 0.0\% [$W=8$]\\
Balanced & H{\"u}sler--Reiss & Hard & \num{500} & 0.0\% & 0.0\% & 0.8\% & 0.0\% [$W=5$]\\
Balanced & H{\"u}sler--Reiss & Hard & \num{750} & 3.0\% & 0.6\% & 5.0\% & 0.0\% [$W=5$]\\
Balanced & H{\"u}sler--Reiss & Hard & \num{1000} & 10.0\% & 2.4\% & 15.2\% & 0.0\% [$W=5$]\\
Unbalanced & Gumbel & Easy & \num{250} & 83.2\% & 77.0\% & 97.2\% & 32.0\% [$W=8$]\\
\addlinespace
Unbalanced & Gumbel & Easy & \num{500} & 98.8\% & 98.8\% & 100.0\% & 92.6\% [$W=5$]\\
Unbalanced & Gumbel & Easy & \num{750} & 100.0\% & 100.0\% & 100.0\% & 97.6\% [$W=5$]\\
Unbalanced & Gumbel & Easy & \num{1000} & 100.0\% & 100.0\% & 100.0\% & 100.0\% [$W=5$]\\
Unbalanced & Gumbel & Moderate & \num{250} & 36.8\% & 27.4\% & 67.6\% & 3.0\% [$W=8$]\\
Unbalanced & Gumbel & Moderate & \num{500} & 88.2\% & 85.0\% & 97.4\% & 40.6\% [$W=10$]\\
\addlinespace
Unbalanced & Gumbel & Moderate & \num{750} & 98.0\% & 97.8\% & 99.4\% & 80.2\% [$W=10$]\\
Unbalanced & Gumbel & Moderate & \num{1000} & 100.0\% & 99.2\% & 99.8\% & 96.8\% [$W=10$]\\
Unbalanced & Gumbel & Hard & \num{250} & 0.0\% & 0.0\% & 0.0\% & 0.0\% [$W=5$]\\
Unbalanced & Gumbel & Hard & \num{500} & 1.0\% & 0.4\% & 5.6\% & 0.0\% [$W=8$]\\
Unbalanced & Gumbel & Hard & \num{750} & 7.6\% & 2.8\% & 27.8\% & 0.4\% [$W=10$]\\
\addlinespace
Unbalanced & Gumbel & Hard & \num{1000} & 22.2\% & 15.4\% & 50.8\% & 4.6\% [$W=8$]\\
Unbalanced & H{\"u}sler--Reiss & Easy & \num{250} & 10.0\% & 0.4\% & 13.2\% & 0.2\% [$W=5$]\\
Unbalanced & H{\"u}sler--Reiss & Easy & \num{500} & 33.4\% & 6.2\% & 43.8\% & 0.6\% [$W=8$]\\
Unbalanced & H{\"u}sler--Reiss & Easy & \num{750} & 59.4\% & 22.4\% & 69.0\% & 2.2\% [$W=10$]\\
Unbalanced & H{\"u}sler--Reiss & Easy & \num{1000} & 80.4\% & 44.4\% & 84.6\% & 5.4\% [$W=15$]\\
\addlinespace
Unbalanced & H{\"u}sler--Reiss & Moderate & \num{250} & 4.4\% & 0.2\% & 8.0\% & 0.0\% [$W=5$]\\
Unbalanced & H{\"u}sler--Reiss & Moderate & \num{500} & 31.0\% & 5.2\% & 40.6\% & 0.0\% [$W=8$]\\
Unbalanced & H{\"u}sler--Reiss & Moderate & \num{750} & 52.8\% & 20.8\% & 69.0\% & 2.2\% [$W=8$]\\
Unbalanced & H{\"u}sler--Reiss & Moderate & \num{1000} & 80.2\% & 41.4\% & 85.4\% & 4.8\% [$W=15$]\\
Unbalanced & H{\"u}sler--Reiss & Hard & \num{250} & 0.0\% & 0.0\% & 0.0\% & 0.0\% [$W=5$]\\
\addlinespace
Unbalanced & H{\"u}sler--Reiss & Hard & \num{500} & 0.0\% & 0.0\% & 0.8\% & 0.0\% [$W=5$]\\
Unbalanced & H{\"u}sler--Reiss & Hard & \num{750} & 1.6\% & 0.0\% & 6.4\% & 0.0\% [$W=8$]\\
Unbalanced & H{\"u}sler--Reiss & Hard & \num{1000} & 7.8\% & 0.0\% & 15.4\% & 0.0\% [$W=8$]\\*
\end{longtable}
\endgroup{}

\subsection{Detailed Results for Simulation Study 2}
\label{sec:supp_simulation2_tables}

Table~\ref{tab:simulation2_mean_ari} reports the mean adjusted Rand
indices and their Monte Carlo standard errors. Table~
\ref{tab:simulation2_exact_recovery} reports the percentage of
replications in which the latent partition is recovered exactly.

\begin{table}
\centering
\caption{\label{tab:simulation2_mean_ari}Mean adjusted Rand index for Simulation Study 2, with Monte Carlo standard errors in parentheses, based on up to 500 replications per setting.}
\centering
\fontsize{8}{10}\selectfont
\begin{tabular}[t]{crrr}
\toprule
$\chi$ & $T$ & Wasserstein--Aitchison & Conditional extremes\\
\midrule
0.1 & \num{10000} & 0.038 (0.003) & 0.024 (0.003)\\
0.1 & \num{100000} & 0.397 (0.005) & 0.279 (0.005)\\
0.3 & \num{10000} & 0.210 (0.005) & 0.140 (0.005)\\
0.3 & \num{100000} & 0.957 (0.003) & 0.384 (0.013)\\
0.5 & \num{10000} & 0.367 (0.006) & 0.217 (0.005)\\
\addlinespace
0.5 & \num{100000} & 0.994 (0.001) & 0.027 (0.003)\\
0.7 & \num{10000} & 0.394 (0.006) & 0.084 (0.005)\\
0.7 & \num{100000} & 0.990 (0.002) & 0.004 (0.001)\\
0.9 & \num{10000} & 0.299 (0.005) & 0.223 (0.005)\\
0.9 & \num{100000} & 0.790 (0.007) & 0.411 (0.009)\\
\bottomrule
\end{tabular}
\end{table}

\begin{table}
\centering
\caption{\label{tab:simulation2_exact_recovery}Percentage of replications achieving exact recovery, $\operatorname{ARI}=1$, in Simulation Study 2.}
\centering
\fontsize{8}{10}\selectfont
\begin{tabular}[t]{crrr}
\toprule
$\chi$ & $T$ & Wasserstein--Aitchison & Conditional extremes\\
\midrule
0.1 & \num{10000} & 0.0\% & 0.0\%\\
0.1 & \num{100000} & 0.0\% & 0.0\%\\
0.3 & \num{10000} & 0.0\% & 0.0\%\\
0.3 & \num{100000} & 66.8\% & 2.2\%\\
0.5 & \num{10000} & 0.0\% & 0.0\%\\
\addlinespace
0.5 & \num{100000} & 95.2\% & 0.0\%\\
0.7 & \num{10000} & 0.0\% & 0.0\%\\
0.7 & \num{100000} & 92.0\% & 0.0\%\\
0.9 & \num{10000} & 0.0\% & 0.0\%\\
0.9 & \num{100000} & 21.6\% & 0.2\%\\
\bottomrule
\end{tabular}
\end{table}

\subsection{Detailed Results for Simulation Study 3}
\label{sec:supp_simulation3_tables}

Table~\ref{tab:simulation3_mean_ari} reports the mean adjusted Rand
indices and their Monte Carlo standard errors. Table~
\ref{tab:simulation3_exact_recovery} reports the percentage of
replications in which the latent partition is recovered exactly.

\begin{table}
\centering
\caption{\label{tab:simulation3_mean_ari}Mean adjusted Rand index for Simulation Study 3, with Monte Carlo standard errors in parentheses, based on 500 independent replications for each value of $\delta$.}
\centering
\fontsize{8}{10}\selectfont
\begin{tabular}[t]{crrr}
\toprule
$\delta$ & Wasserstein--Aitchison & Wasserstein--simplex & Conditional extremes\\
\midrule
0.20 & 0.069 (0.006) & 0.085 (0.007) & 0.055 (0.006)\\
0.40 & 0.336 (0.009) & 0.351 (0.009) & 0.229 (0.009)\\
0.60 & 0.598 (0.011) & 0.624 (0.011) & 0.466 (0.011)\\
0.80 & 0.867 (0.008) & 0.881 (0.007) & 0.745 (0.010)\\
1.00 & 0.977 (0.003) & 0.981 (0.003) & 0.939 (0.005)\\
\bottomrule
\end{tabular}
\end{table}

\begin{table}
\centering
\caption{\label{tab:simulation3_exact_recovery}Percentage of the 500 replications achieving exact recovery, $\operatorname{ARI}=1$, in Simulation Study 3.}
\centering
\fontsize{8}{10}\selectfont
\begin{tabular}[t]{crrr}
\toprule
$\delta$ & Wasserstein--Aitchison & Wasserstein--simplex & Conditional extremes\\
\midrule
0.20 & 0.0\% & 0.0\% & 0.0\%\\
0.40 & 0.8\% & 1.0\% & 0.0\%\\
0.60 & 15.6\% & 18.2\% & 6.2\%\\
0.80 & 59.2\% & 62.8\% & 35.0\%\\
1.00 & 91.4\% & 93.0\% & 78.8\%\\
\bottomrule
\end{tabular}
\end{table}

\section{Construction of the trivariate angular measures}\label{sec:trivariate_construction}

This Section describes in detail the construction of the three trivariate angular measures used in the trivariate simulation study, together with the generation of the corresponding observations.

\subsection*{Bernstein--Dirichlet representation}

Let $\Delta_2^\circ = \{ \theta = (\theta_1,\theta_2,\theta_3) \in (0,1)^3 : \theta_1+\theta_2+\theta_3=1 \}$ denote the interior of the two-dimensional simplex.

Finite mixtures of Dirichlet densities provide a flexible representation of angular measures in arbitrary dimensions \citep{boldi2007mixture,sabourin2014bayesian}. Our construction uses the connection between integer-valued Dirichlet densities and multivariate Bernstein polynomial bases \citep{marcon2016bayesian,marcon2017multivariate}. We fix an integer order $J=18$ and define
\[
\mathcal{I}_J = \{ \alpha = (\alpha_1,\alpha_2,\alpha_3) \in \mathbb{N}^3 : \alpha_j \ge 1 \; (j=1,2,3), \; \alpha_1+\alpha_2+\alpha_3=J \}.
\]
Since every component of $\alpha$ is strictly positive, all Dirichlet components are supported on the open simplex. The number of basis components is $|\mathcal{I}_J| = \binom{J-1}{2} = 136$. For each $\alpha \in \mathcal{I}_J$, define
\[
b_{\alpha}(\theta) = \frac{\Gamma(J)}{\Gamma(\alpha_1)\Gamma(\alpha_2)\Gamma(\alpha_3)} \prod_{j=1}^{3} \theta_j^{\alpha_j-1} \qquad (\theta \in \Delta_2^\circ),
\]
which is the density of a Dirichlet distribution with parameter $\alpha$. Given non-negative weights $q = (q_{\alpha})_{\alpha \in \mathcal{I}_J}$ with $\sum_{\alpha \in \mathcal{I}_J} q_{\alpha} = 1$, we define the angular density
\[
h_{q}(\theta) = \sum_{\alpha \in \mathcal{I}_J} q_{\alpha} b_{\alpha}(\theta).
\]
The corresponding angular probability measure is denoted by $H_{q}$.

\subsection*{Angular moment constraints}

Under the probability normalization of the angular measure, a valid three-dimensional angular measure must satisfy
\[
\int_{\Delta_2^\circ} \theta_j \, \mathrm{d}H_{q}(\theta) = \frac{1}{3} \qquad (j=1,2,3).
\]
For a Dirichlet random vector with parameter $\alpha$, $\mathbb{E}_{\alpha}(\Theta_j) = \alpha_j / J$. The moment constraints can therefore be written as linear constraints on the mixture weights:
\[
\sum_{\alpha \in \mathcal{I}_J} q_{\alpha} \frac{\alpha_j}{J} = \frac{1}{3} \qquad (j=1,2,3).
\]
Since $\Theta_1+\Theta_2+\Theta_3=1$, the third constraint is implied by the first two together with the unit-mass constraint. Consequently, only the constraints for $j=1,2$ are supplied to the numerical linear program.

\subsection*{Fixing the pairwise tail-dependence coefficients}

Let $\Theta = (\Theta_1,\Theta_2,\Theta_3) \sim H_{q}$. Under the probability normalization adopted here, $H_{q}$ has total mass one and satisfies $\mathbb{E}_{H_{q}}(\Theta_j) = 1/3$ ($j=1,2,3$). The corresponding spectral measure is therefore $3H_{q}$. It follows that the pairwise tail-dependence coefficient is
\[
\chi_{ij} = 3 \int_{\Delta_2^\circ} \min(\theta_i,\theta_j) \, \mathrm{d}H_{q}(\theta) = 3 \mathbb{E}_{H_{q}}\! \left\{ \min(\Theta_i,\Theta_j) \right\}.
\]
Recall that $H_{q}$ is a mixture of Dirichlet distributions with density $h_{q}(\theta) = \sum_{\alpha \in \mathcal{I}_J} q_{\alpha} b_{\alpha}(\theta)$.To calculate the expectation under this mixture, we first calculate it under each Dirichlet component. For every $\alpha \in \mathcal{I}_J$, let $\Theta^{(\alpha)} = (\Theta_1^{(\alpha)}, \Theta_2^{(\alpha)}, \Theta_3^{(\alpha)}) \sim \operatorname{Dirichlet}(\alpha)$, and define
\[
m_{ij}(\alpha) = \mathbb{E}\! \left[ \min\left( \Theta_i^{(\alpha)}, \Theta_j^{(\alpha)} \right) \right].
\]
This expectation is available analytically from standard properties of the Dirichlet distribution (Kotz et al., 2000, Ch.~49). Let $I_x(a,b) = \operatorname{B}_x(a,b) / \operatorname{B}(a,b)$ denote the regularized incomplete beta function, where $\operatorname{B}_x(a,b) = \int_0^x u^{a-1}(1-u)^{b-1} \, \mathrm{d}u$ and $\operatorname{B}(a,b) = \operatorname{B}_1(a,b)$. If $U \sim \operatorname{Beta}(a,b)$, then $I_x(a,b) = \operatorname{pr}(U \le x)$.

For a fixed component $\alpha$, define $S_{ij}^{(\alpha)} = \Theta_i^{(\alpha)} + \Theta_j^{(\alpha)}$ and $U_{ij}^{(\alpha)} = \Theta_i^{(\alpha)} / (\Theta_i^{(\alpha)} + \Theta_j^{(\alpha)})$. Aggregation properties of the Dirichlet distribution imply that $U_{ij}^{(\alpha)} \sim \operatorname{Beta}(\alpha_i,\alpha_j)$, $\mathbb{E}(S_{ij}^{(\alpha)}) = (\alpha_i+\alpha_j)/J$, and that $S_{ij}^{(\alpha)}$ and $U_{ij}^{(\alpha)}$ are independent. Moreover,
\[
\min\left(\Theta_i^{(\alpha)}, \Theta_j^{(\alpha)}\right) = S_{ij}^{(\alpha)} \min\left(U_{ij}^{(\alpha)}, 1-U_{ij}^{(\alpha)}\right).
\]
Using independence, we obtain
\[
m_{ij}(\alpha) = \mathbb{E}\left(S_{ij}^{(\alpha)}\right) \left[ \mathbb{E}\left\{ U_{ij}^{(\alpha)} I\left(U_{ij}^{(\alpha)} \le \frac{1}{2}\right) \right\} + \mathbb{E}\left\{ \left(1-U_{ij}^{(\alpha)}\right) I\left(U_{ij}^{(\alpha)} > \frac{1}{2}\right) \right\} \right],
\]
where $I(\cdot)$ is the indicator function. The truncated expectations yield
\[
m_{ij}(\alpha) = \frac{1}{J} \left[ \alpha_i I_{1/2}(\alpha_i+1,\alpha_j) + \alpha_j \left\{ 1 - I_{1/2}(\alpha_i,\alpha_j+1) \right\} \right].
\]
Linearity of expectation gives
\[
\chi_{ij} = 3 \sum_{\alpha \in \mathcal{I}_J} q_{\alpha} m_{ij}(\alpha).
\]
The mixture weights are constrained so that $3 \sum_{\alpha \in \mathcal{I}_J} q_{\alpha} m_{ij}(\alpha) = 0.45$ for $(i,j) \in \{(1,2),(1,3),(2,3)\}$. Since $m_{ij}(\alpha)$ is fixed for every $\alpha \in \mathcal{I}_J$, these conditions are linear constraints in $\{q_{\alpha} : \alpha \in \mathcal{I}_J\}$.

\subsection*{Constructing angular measures with different orientations}

To distinguish angular distributions satisfying the same moment and pairwise tail-dependence constraints, we introduce the alternating polynomial
\[
\psi(\theta) = (\theta_1-\theta_2)(\theta_2-\theta_3)(\theta_3-\theta_1) \qquad (\theta \in \Delta_2^\circ).
\]
This polynomial is invariant under cyclic permutations of coordinates and changes sign when any two coordinates are exchanged. We use its expectation as a measure of orientation:
\[
\operatorname{Or}(H) = \int_{\Delta_2^\circ} \psi(\theta) \, \mathrm{d}H(\theta).
\]
For a Dirichlet component with parameter $\alpha$, standard Dirichlet moments yield
\[
o_{\alpha} = \mathbb{E}\left\{ \psi\left(\Theta^{(\alpha)}\right) \right\} = \frac{(\alpha_1-\alpha_2)(\alpha_2-\alpha_3)(\alpha_3-\alpha_1)}{J(J+1)(J+2)}.
\]
The positive-orientation angular measure $H_+$ is obtained by maximizing $\sum_{\alpha \in \mathcal{I}_J} q_{\alpha} o_{\alpha}$ subject to $q_{\alpha} \ge 0$, $\sum_{\alpha \in \mathcal{I}_J} q_{\alpha} = 1$, the first moment constraints, and the pairwise coefficient constraints. The negative-orientation measure $H_-$ minimizes the same objective subject to the same constraints.

Let $\mathcal{O}_+ = \operatorname{Or}(H_+)$ and $\mathcal{O}_- = \operatorname{Or}(H_-)$. Define $\lambda = -\mathcal{O}_- / (\mathcal{O}_+ - \mathcal{O}_-)$ and construct the orientation-balanced measure $H_0 = \lambda H_+ + (1-\lambda) H_-$. By construction, $\operatorname{Or}(H_0) = 0$. Because all constraints are linear in the angular measure, $H_0$, $H_+$, and $H_-$ share the same first angular moments and pairwise tail-dependence coefficients.

\subsection*{Controlling cluster separation with $\delta$}

For $\delta \in [0,1]$, define the three angular measures
\[
H_{+,\delta} = (1-\delta)H_0 + \delta H_+, \qquad H_{-,\delta} = (1-\delta)H_0 + \delta H_-, \qquad H_{0,\delta} = H_0.
\]
Because moment and pairwise constraints are linear, they are preserved under convex mixing for every $g \in \{+,0,-\}$ and $\delta \in [0,1]$. Furthermore, $\operatorname{Or}(H_{+,\delta}) = \delta \mathcal{O}_+$, $\operatorname{Or}(H_{-,\delta}) = \delta \mathcal{O}_-$, and $\operatorname{Or}(H_{0,\delta}) = 0$.

The parameter $\delta$ acts as an interpolation parameter controlling clustering difficulty without altering first moments or pairwise tail-dependence strength:
\begin{itemize}
    \item when $\delta=1$, $H_{+,\delta} = H_+$ and $H_{-,\delta} = H_-$, corresponding to the extreme LP solutions;
    \item as $\delta$ decreases, $H_{+,\delta}$ and $H_{-,\delta}$ move toward $H_0$;
    \item when $\delta=0$, all three measures coincide with $H_0$.
\end{itemize}

\subsection*{Generation of the observations}

Let $g(i) \in \{+,0,-\}$ denote the latent group of random vector $i$. For each $\delta$, angular observations are generated independently as $\Theta_{i,t} \sim H_{g(i),\delta}$. Independently, standard Pareto radial variables $R_{i,t}$ are generated with $\operatorname{pr}(R_{i,t} > r) = 1/r$ ($r \ge 1$). The trivariate observation is defined as $Z_{i,t} = 3 R_{i,t} \Theta_{i,t}$.

Since $\mathbb{E}(\Theta_{i,t}^{(j)}) = 1/3$, the components of $Z_{i,t}$ have unit-Pareto upper tails. Specifically, for $z \ge 3$,
\[
\operatorname{pr}\left(Z_{i,t}^{(j)} > z\right) = \mathbb{E}\left\{ \operatorname{pr}\left( R_{i,t} > \frac{z}{3\Theta_{i,t}^{(j)}} \;\middle|\; \Theta_{i,t} \right) \right\} = \frac{3}{z} \mathbb{E}\left(\Theta_{i,t}^{(j)}\right) = \frac{1}{z}.
\]
Furthermore, $\|Z_{i,t}\|_1 = 3 R_{i,t}$ and $Z_{i,t} / \|Z_{i,t}\|_1 = \Theta_{i,t}$, so the angular component of every generated observation follows exactly $H_{g(i),\delta}$.

\section{Supplementary Results for the Data Applications}
\label{sec:supplementary_applications}

\subsection{Additional Results for the International Equity Markets}
\label{sec:supplementary_finance}

\begin{figure}
    \centering
    \includegraphics[width=\textwidth]{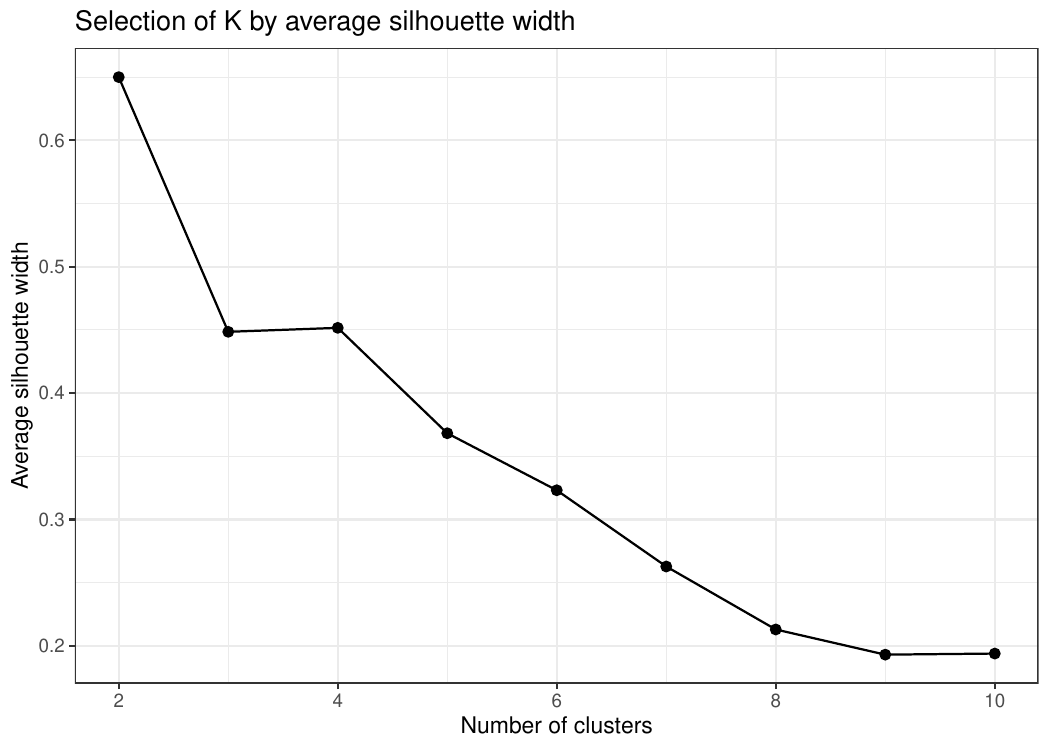}
    \caption{Average silhouette width as a function of the number of clusters. The criterion is maximized at $K=2$.}
    \label{fig:silhouette_selection_finance}
\end{figure}

\begin{figure}
    \centering
    \includegraphics[width=\textwidth]{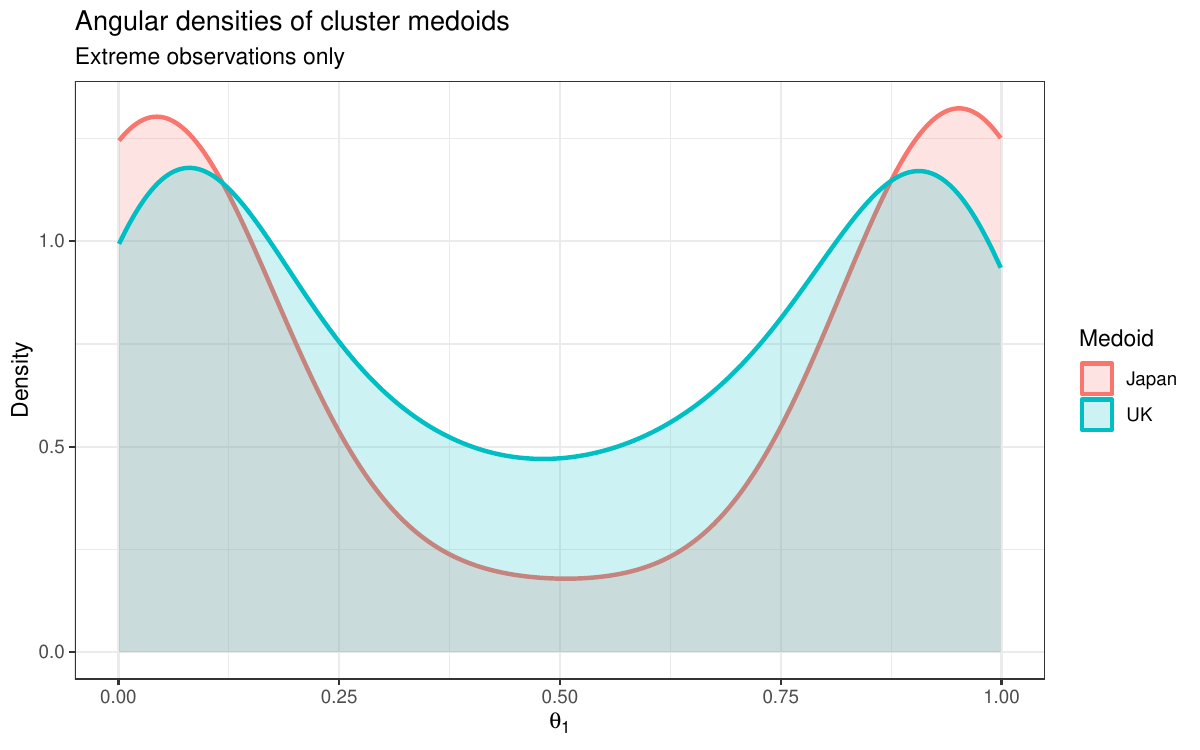}
    \caption{Kernel density estimates of the angular distributions associated with the two cluster medoids. The variable $\theta_1$ denotes the angular share corresponding to the United States market.}
    \label{fig:finance_medoids}
\end{figure}

To examine the extremal-dependence regimes represented by the two cluster medoids, Fig.~\ref{fig:finance_medoids} displays kernel density estimates of their angular distributions. In the bivariate setting, the angular component is completely characterized by $\theta_1$, which represents the proportion of an extreme event attributable to the United States market. Values close to one indicate events dominated by United States losses, whereas values close to zero indicate events dominated by local-market losses. The United Kingdom medoid exhibits more mass near the centre of the simplex, indicating a relatively balanced sharing of extreme losses between the United States and United Kingdom markets. In contrast, the Japan medoid displays a more pronounced concentration near the boundaries, suggesting that extreme events are more frequently dominated by one market or the other.

\begin{figure}
    \centering
    \includegraphics[width=0.85\textwidth]{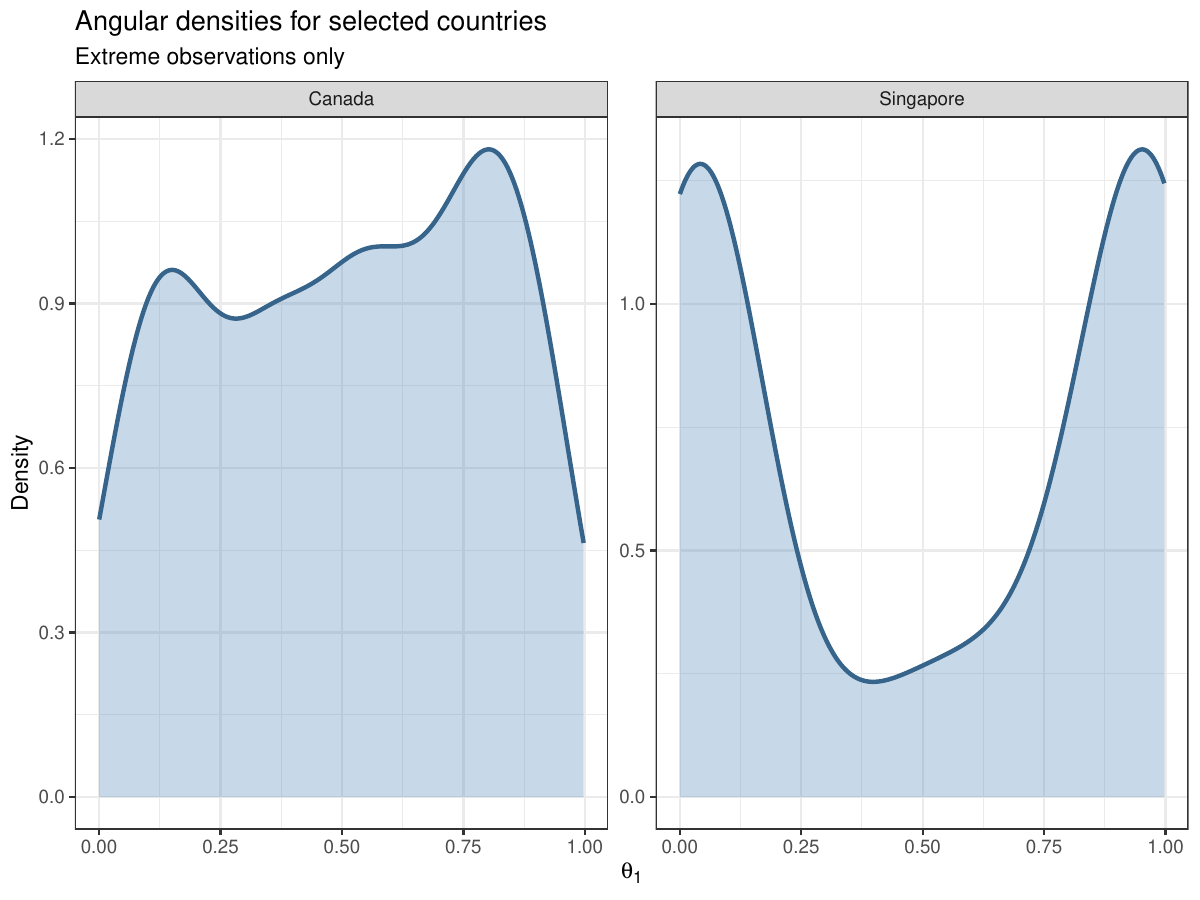}
    \caption{Estimated angular densities for selected countries located near the boundary of the two clusters.}
    \label{fig:finance_boundary_countries}
\end{figure}

We next examine several countries located near the boundary of the two clusters. Figure~\ref{fig:finance_boundary_countries} displays the estimated angular densities for Canada and Singapour. Canada exhibit substantially more mass near the centre of the simplex. This indicates a strong extremal dependence between United States and Canada losses during extreme events than that observed for the United Kingdom medoid. In contrast, Singapour shows much weaker dependence with US market even weaker than the Japan medoid and occupies a peripheral position in the multidimensional scaling representation shown in Fig.~\ref{fig:finance_clusters}. These observations suggest that Saudi Arabia lies near the boundary between the two identified extremal-dependence regimes.

\begin{figure}
    \centering
    \includegraphics[width=0.85\textwidth]{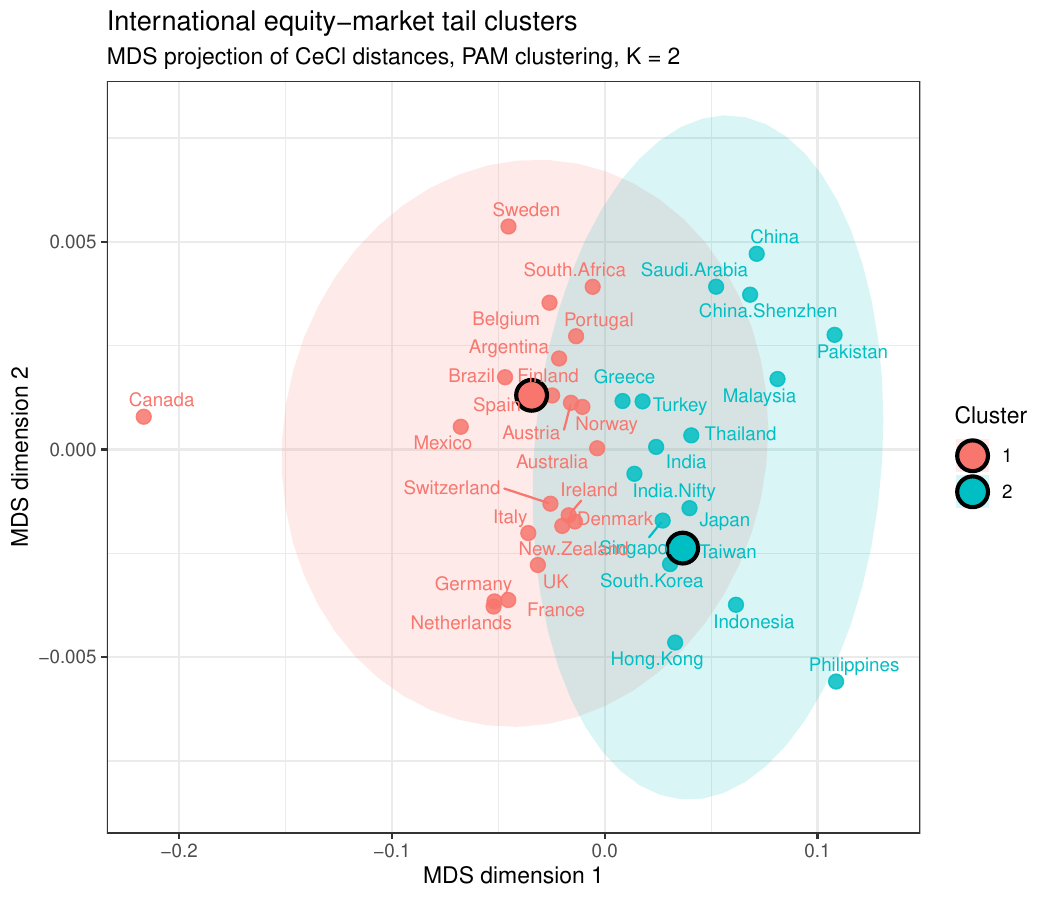}
    \caption{International equity-market tail clusters obtained with the conditional extremes clustering dissimilarity. The figure shows a two-dimensional multidimensional scaling projection of the dissimilarities together with the PAM partition. Black-circled observations indicate the cluster medoids.}
    \label{fig:cecl_country_clusters}
\end{figure}

As a robustness comparison, we applied the conditional extremes clustering methodology of \citet{o2025clustering} to the same data set. It yields exactly the same two-cluster partition as the Wasserstein--Aitchison procedure. Figure~\ref{fig:cecl_country_clusters} shows the corresponding multidimensional scaling representation. This agreement indicates that the identified grouping is not an artefact of the particular dissimilarity measure employed, but reflects a structure in the extremal dependence of the international equity markets.

\subsection{Additional Details for the Swiss Catchment Application}
\label{sec:supplementary_swiss}

We use the observation-based time-series collection from version 0.8 of CAMELS-CH \citep{hoge2023camelsch,hoge2023camelschdata}. The complete data set contains daily hydrometeorological observations and catchment attributes for 331 hydrological units, comprising 298 river catchments and 33 lakes. The domain covers Switzerland and Liechtenstein together with transboundary catchments in Austria, France, Germany and Italy.

For catchment $i$ and day $t$, the original variables are catchment-mean precipitation $p_{i,t}$ in mm\,d$^{-1}$, catchment-mean daily air temperature $T_{i,t}$ in $^\circ$C, snow water equivalent $s_{i,t}$ in mm, and observed specific discharge $q_{i,t}$ in mm\,d$^{-1}$. Precipitation and temperature originate from the 2-km observation-based products of MeteoSwiss and were spatially averaged over each catchment. Discharge was recorded at the corresponding gauging stations and converted to specific discharge. The snow-water-equivalent data were obtained by aggregating a distributed 1-km snow-cover product constrained by observations from 345 snow-monitoring stations and MeteoSwiss precipitation data; the associated data-assimilation methodology is described by \citet{magnusson2014swe}. Snow water equivalent should therefore be regarded as a model--data-assimilation product rather than a direct catchment-scale measurement. Its availability from September 1998 determines the analysis period of 1 September 1998--31 December 2020. Daily precipitation is partitioned into liquid and solid fractions using the temperature-dependent liquid fraction
\begin{equation}
f(T)
=
\begin{cases}
0,   & T\leq 0\,^\circ\mathrm{C},\\[2mm]
T/2, & 0\,^\circ\mathrm{C}<T<2\,^\circ\mathrm{C},\\[2mm]
1,   & T\geq 2\,^\circ\mathrm{C}.
\end{cases}
\label{eq:rain_fraction}
\end{equation}
Following the temperature-range approach discussed by \citet{jennings2019phase},
we use a simple linear interpolation between \(0\) and \(2\,^\circ\mathrm{C}\),
centred on their Northern Hemisphere mean 50\% rain--snow threshold of
approximately \(1\,^\circ\mathrm{C}\).

The three event-scale variables are defined by
\begin{align}
L_{i,t}
&=
\sum_{j=0}^{2}
p_{i,t-j}f\left(T_{i,t-j}\right),
\label{eq:liquid_rain}\\
M_{i,t}
&=
\left(s_{i,t-3}-s_{i,t}\right)_{+},
\qquad
(x)_{+}=\max(x,0),
\label{eq:swe_depletion}\\
Q^{\star}_{i,t}
&=
\max_{h\in\{0,1,2\}}q_{i,t+h}.
\label{eq:runoff_response}
\end{align}
Thus, $L_{i,t}$ is accumulated liquid precipitation over the current and two preceding days, $M_{i,t}$ is positive net snow-water-equivalent depletion over three days, and $Q^{\star}_{i,t}$ is the maximum specific discharge on day $t$ or either of the following two days.

The three-day event scale is motivated by the rapid response of many Alpine catchments. \citet{froidevaux2015flood} found that precipitation during the days immediately preceding a flood was generally the most relevant for Swiss Alpine catchments. Three-day precipitation totals and snow-water-equivalent decreases have also been used to distinguish rainfall-driven, snowmelt-driven and mixed rainfall--snowmelt runoff events \citep{davenport2020flood}. The chosen window is nevertheless an approximation and should not be interpreted as a universal catchment-response time.
Net snow-water-equivalent depletion is a proxy for snowpack water release rather than gross snowmelt. Melt occurring simultaneously with new snowfall may be partly or completely masked by increasing snow water equivalent. Negative depletion is therefore set to zero rather than interpreted as negative melt.

Negative precipitation, snow water equivalent and discharge values were treated as missing, and missing snow-water-equivalent observations were not replaced by zero. A daily triplet
\[
X_{i,t}
=
\left(
L_{i,t},
M_{i,t},
Q^{\star}_{i,t}
\right)
\]
was retained only when all observations required by equations~\eqref{eq:liquid_rain}--\eqref{eq:runoff_response} were available. Catchments with fewer than \num{3650} complete triplets, corresponding approximately to ten years of daily observations, were excluded. Selection was based exclusively on data availability.

\begin{figure}
    \centering
    \includegraphics[width=0.82\textwidth]
    {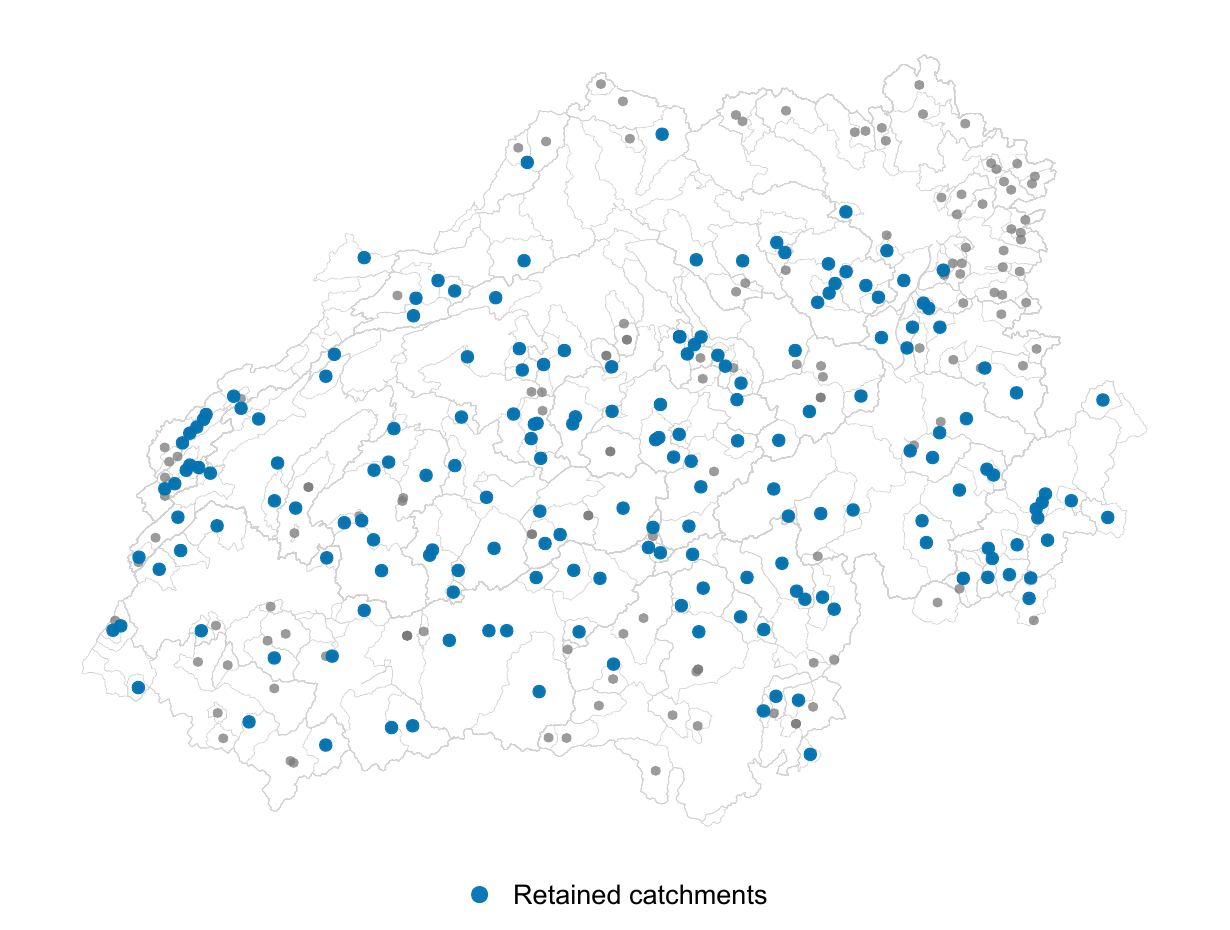}
    \caption{Locations of the 331 catchments in the CAMELS-CH data set. Blue points indicate the 182 catchments retained for analysis. Grey points indicate catchments excluded because they had fewer than 10 years of available data or fewer than 50 declustered extreme events.}
    \label{fig:camels_retained_catchments}
\end{figure}

Within each retained catchment, the three margins were transformed separately to standard Pareto scale. Because $L_{i,t}$ and $M_{i,t}$ contain substantial point masses at zero, tied observations were assigned reproducible random ranks within their tie blocks. For marginal sample size $n_i$, we used
\[
U_{i,t,j}
=
\frac{\operatorname{rank}(X_{i,t,j})}{n_i+1},
\qquad
Z_{i,t,j}
=
\frac{1}{1-U_{i,t,j}},
\qquad
j=1,2,3.
\]
The radial component was then
\[
R_{i,t}
=
\sum_{j=1}^{3}Z_{i,t,j}.
\]
Observations above the catchment-specific 95th percentile of $R_{i,t}$ were identified as multivariate extremes. Successive exceedances separated by at most three days were assigned to the same event, of which only the observation with the largest radial component was retained. This is our de-clustering method similarly as what was done by \citet{sabourin2015combining}. Requiring at least 50 such events left 182 of the original 331 catchments, whose locations are shown in Fig.~\ref{fig:camels_retained_catchments}.

\begin{figure}
    \centering
    \includegraphics[width=0.85\textwidth]
    {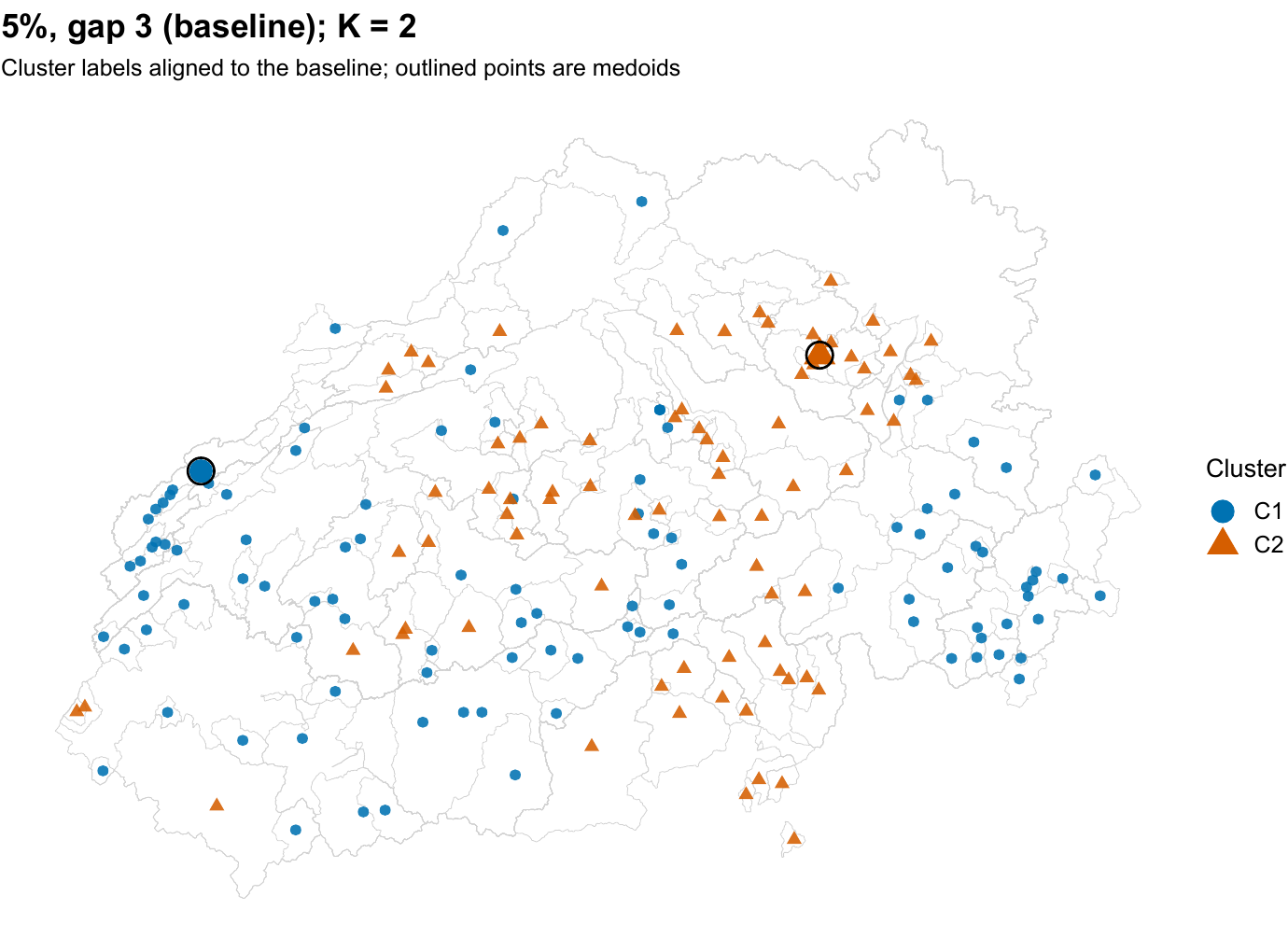}
    \caption{Spatial distribution of the two extremal-dependence clusters. Blue circles, orange triangles denote C1, C2, respectively. Larger outlined symbols identify the cluster medoids. Light-grey lines delimit the catchments included in the analysis.}
    \label{fig:camelsch_map}
\end{figure}

\begin{figure}
    \centering
    \includegraphics[width=0.82\textwidth]
    {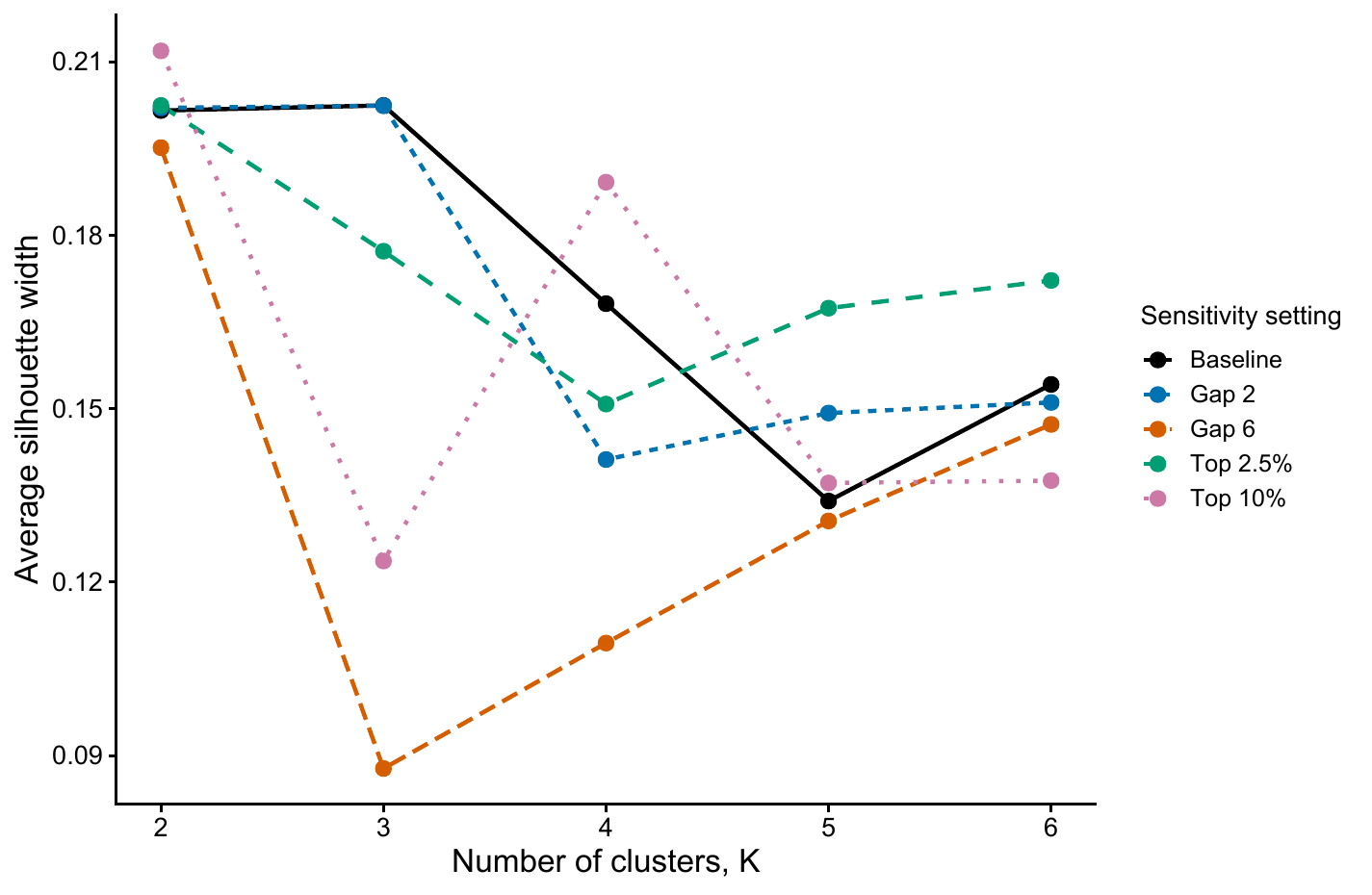}
    \caption{Average silhouette width as a function of the number of clusters under the baseline and alternative radial-threshold and declustering-window specifications. Although some specifications marginally favour three clusters, the evidence is not consistent across settings and generally supports a two-cluster representation.}
    \label{fig:camels_sensitivity_silhouette}
\end{figure}

The corresponding angular observations were transformed to two-dimensional Aitchison coordinates. Pairwise 1-Wasserstein distances were computed between the empirical Aitchison-coordinate distributions and clustered using partitioning around medoids. We considered $K=2,\ldots,6$; the average silhouette width shown in Fig.~\ref{fig:camels_sensitivity_silhouette} selected $K=3$. The resulting partition exhibits broad spatial organization without forming strictly contiguous regions, as shown in Fig.~\ref{fig:camelsch_map}

The interpretation of the two clusters was assessed using four CAMELS-CH attributes that were not used in the clustering: mean catchment elevation, fraction of precipitation falling as snow, mean catchment slope and mean half-flow date. The half-flow date is the number of days from the beginning of the hydrological year, defined as 1 October, required for cumulative discharge to reach one half of annual discharge \citep{hoge2023camelsch}. Figure~\ref{fig:camelsch_validation} shows that C1 catchments have the highest mean elevations and largest fractions of precipitation falling as snow. They also have steeper slopes and later half-flow dates, consistently with delayed runoff associated with seasonal snow storage and melt. C2 catchments have the lower elevations and snow fractions, supporting their interpretation as predominantly rainfall--runoff catchments. This is consistent with the angular regime identified in Fig.~\ref{fig:camelsch_simplex} of the main paper.
% The two clusters differ significantly in mean elevation, the fraction of precipitation falling as snow, and mean half-flow date. Together with the angular geometry and differences in runoff timing, these results support interpreting one cluster as more strongly influenced by snowmelt an rainfall, and the other as predominantly rainfall-driven.
% \begin{figure}
%     \centering
%     \includegraphics[width=\textwidth]
%     {art/physical_validation_baseline_K2.pdf}
%     \caption{External physical validation of the two extremal-dependence clusters. The panels show mean catchment elevation, fraction of precipitation falling as snow, mean catchment slope and mean half-flow date. Individual symbols represent catchments, with circles, triangles corresponding to C1, C2. Lowercase letters denote significance groups based on pairwise Wilcoxon rank-sum tests at the 5\% level.}
%     \label{fig:camelsch_validation}
% \end{figure}
We assessed sensitivity to the declustering window and radial threshold by comparing the baseline specification (upper $5\%$ of radial observations and a three-day window) with windows of two and six days and retained fractions of $2.5\%$ and $10\%$. The average silhouette widths in Fig.~\ref{fig:camels_sensitivity_silhouette} generally support two clusters; although $K=3$ is marginally favoured in some specifications, this preference is not consistent. Table~\ref{tab:camels_sensitivity_agreement} compares each two-cluster partition with the baseline. After aligning cluster labels, $89.0\%$--$99.5\%$ of catchments retain their assignment, with adjusted Rand indices between $0.607$ and $0.978$. The partition is nearly invariant to the declustering window and moderately more sensitive to the threshold, consistent with the usual bias--variance trade-off in extreme-value analysis. Overall, these results support $K=2$ as the most stable choice.
\begin{table}
\centering
\caption{Sensitivity of the clustering results to the radial threshold and
declustering window. Each partition is compared with the baseline specification
(upper $5\%$ and a three-day window) using the 182 catchments common to all
specifications. Direct agreement is computed after optimal alignment of the
cluster labels. ARI denotes the adjusted Rand index.}
\label{tab:camels_sensitivity_agreement}
\small
\setlength{\tabcolsep}{4pt}
\begin{tabular}{clrrrrr}
\hline
$K$ &
Sensitivity scenario &
\shortstack{Common\\catchments} &
\shortstack{Same\\assignment} &
\shortstack{Changed\\assignment} &
\shortstack{Direct\\agreement} &
ARI \\
\hline
2 & $5\%$, gap 2        & 182 & 179 &  3 & $98.4\%$ & 0.935 \\
2 & $5\%$, gap 6        & 182 & 181 &  1 & $99.5\%$ & 0.978 \\
2 & Top $2.5\%$, gap 3  & 182 & 162 & 20 & $89.0\%$ & 0.607 \\
2 & Top $10\%$, gap 3   & 182 & 167 & 15 & $91.8\%$ & 0.695 \\
\hline
3 & $5\%$, gap 2        & 182 & 178 &  4 & $97.8\%$ & 0.927 \\
3 & $5\%$, gap 6        & 182 &  99 & 83 & $54.4\%$ & 0.307 \\
3 & Top $2.5\%$, gap 3  & 182 & 126 & 56 & $69.2\%$ & 0.363 \\
3 & Top $10\%$, gap 3   & 182 & 101 & 81 & $55.5\%$ & 0.336 \\
\hline
4 & $5\%$, gap 2        & 182 & 115 & 67 & $63.2\%$ & 0.442 \\
4 & $5\%$, gap 6        & 182 & 114 & 68 & $62.6\%$ & 0.335 \\
4 & Top $2.5\%$, gap 3  & 182 & 119 & 63 & $65.4\%$ & 0.326 \\
4 & Top $10\%$, gap 3   & 182 & 161 & 21 & $88.5\%$ & 0.709 \\
\hline
\end{tabular}
\end{table}

% =============================================================================
% References (style set by the sn-mathphys-ay class option)
% =============================================================================

\clearpage % flush remaining supplementary floats before the reference list
\bibliography{project}

\end{document}